\documentclass[preprint,12pt]{elsarticle}

\usepackage{lineno}

\usepackage{epsfig}
\usepackage{array,tabularx,epsfig,mathrsfs,graphicx,rotating}
\usepackage{ifthen}
\usepackage{ragged2e}
\PassOptionsToPackage{hyphens}{url}
\usepackage[hyphens]{url}
\usepackage{hyperref}
\usepackage{listings}
\usepackage{epstopdf}
\usepackage{color}
\usepackage{float}
\usepackage{subfig}

\usepackage{lineno}
\modulolinenumbers[5]
\usepackage{ulem}
\usepackage{amssymb}
\usepackage{booktabs}
\usepackage{graphicx}
\usepackage{physics}
\usepackage{dcolumn}
\usepackage{bm}
\usepackage{xcolor}

\hypersetup{
  colorlinks=true,
  linkcolor=blue,
  citecolor=blue,
  urlcolor=blue
}

\journal{Nuclear Physics A}

\begin{document}

\begin{frontmatter}

\title{
On the resolution of dual readout calorimeters}
\author[a]{S.~Eno}
\author[b]{L.~Wu}

\author[a]{M.~Y.~Aamir}

\author[c]{S.V.~Chekanov}
\author[a]{S.~Nabili}
\author[a]{C.~Palmer}


\address[a]{Department of  Physics, U. Maryland,
             College Park,
             Maryland,
             MD 20742,
             USA}

\address[b]{Stanford University, Stanford, CA 94305, USA
    }

\address[c]{HEP Division,
             Argonne National Laboratory, 9700 S.~Cass Avenue,
             Lemont,
             60439,
             IL,
             USA}



\begin{abstract}
Dual readout calorimeters allow state-of-the-art resolutions for hadronic energy measurements.  Their various incarnations are leading candidates for the calorimeter systems for future colliders.  In this paper, we present a simple formula for the resolution of a dual readout calorimeter, which we verify with a toy simulation and with full simulation results.  This formula can help those new to dual readout calorimetry understand its strengths and limitations.  The paper also highlights that the dual readout correction works not just to compensate for binding energy loss, but also for energies escaping the calorimeter or clustering algorithm.  Formulae are also presented for approximate resolutions and energy scales in terms of different sources of response.

\end{abstract}


\begin{keyword}

Calorimeters, Detector modelling and simulations



\end{keyword}

\end{frontmatter}

\section{\label{sec:intro}Introduction}
Because they allow state-of-the-art resolutions for hadronic energy measurements, dual readout calorimeters~\cite{Lee:2017xss}  are leading candidates for the calorimetry systems of future collider detectors~\cite{ALY2020162088,Lucchini:2020bac}.  In this paper, we present a simple formula predicting the resolution of a dual readout calorimeter, and verify it using both toy  and full simulations.
The formula gives insight into the strengths and limitations of Dual Readout Calorimetry, and its underlying physics.   Formulae are also presented for the resolutions and energy scales in terms of different sources of response.

\begin{figure}[hbtp]
\centering
\resizebox{0.49\textwidth}{!}{\includegraphics{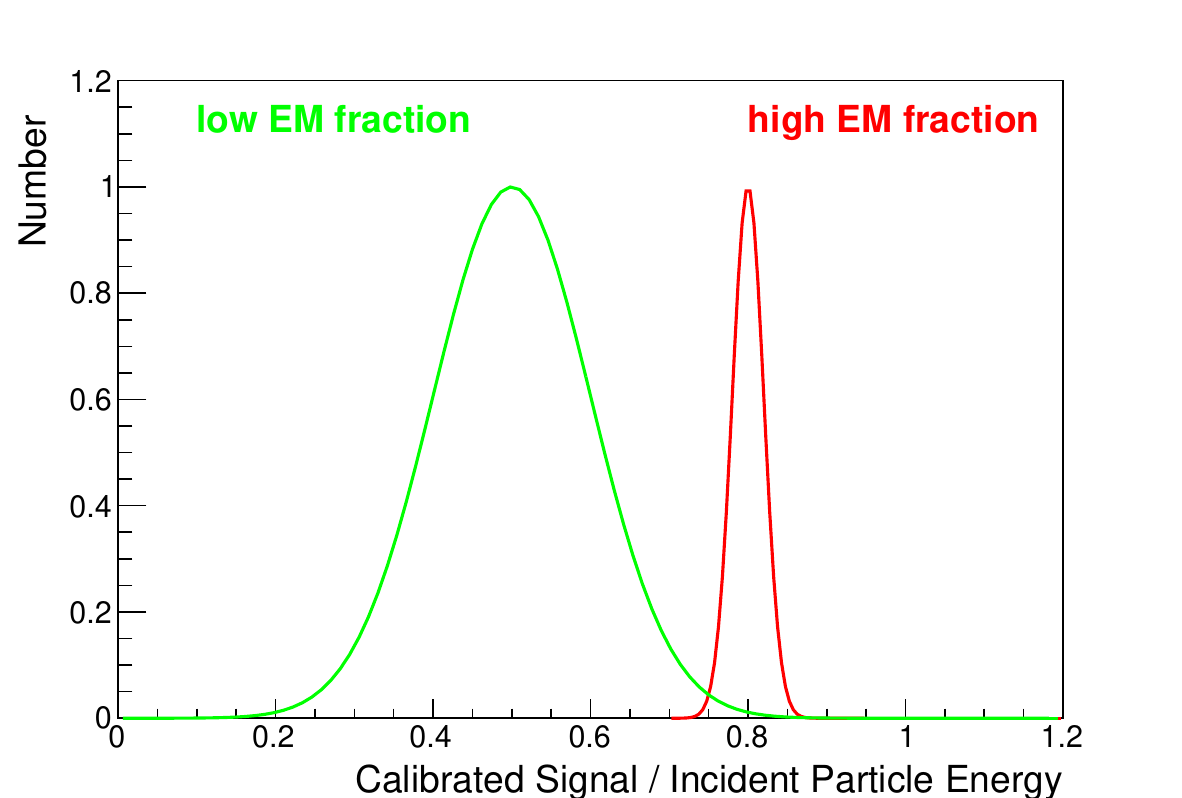}}
\resizebox{0.44\textwidth}{!}{\includegraphics{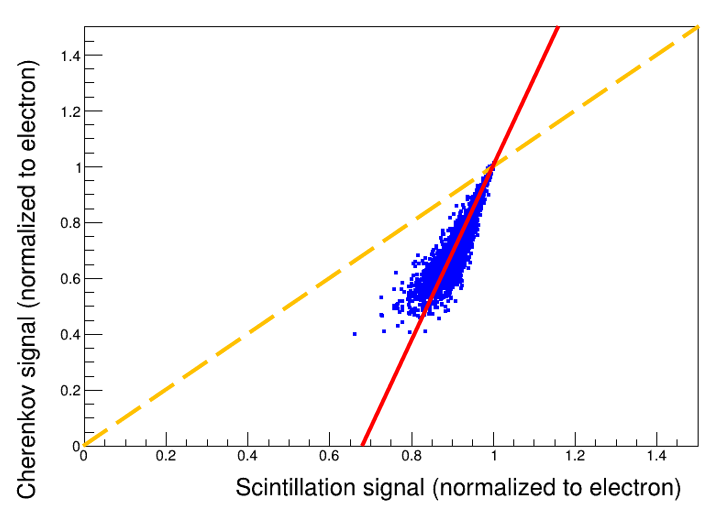}}
\caption{ [left]
Inspired by Ref.~\cite{Lee:2017xss}, an illustration of how fluctuations in the subcomponents of a shower can dominate the resolutions of hadronic calorimeters. The two Gaussians represent the typical measured signal for a fixed incident particle energy for two different values of the electromagnetic fraction.
[right] An illustration of how two measurements, one with sensitivity to all components of the shower (``S'') and another with reduced sensitivity to the  portion of the shower not from photons (``C'') can be used to correct for fluctuations in the energy scale due to fluctuations in the subcomponent composition.  The points represent a pair of ``C'' and ``S'' measurements from an individual hadronic shower as introduced in Eqs.~\ref{eqn:eq1}-\ref{eqn:eq2}.  The label ``EM objects" refers to photons and electrons.
The dual readout correction moves points parallel to the red line to their intersection with the  orange dashed line, reducing their spread.
}\label{fig:wigmans1}
\end{figure}

While calorimeters that have  excellent resolutions for electrons and photons have existed for a long time, achieving excellent resolutions for hadronic particles is challenging. In hadronic showers, the breakup of nuclei requires overcoming their binding energy, an energy-loss mechanism that does not produce detectable signals. The amount of lost energy is correlated with the number of nuclear breakups.
Nuclear breakups can also lead to other sources of lost energy. The released neutrons, due to their low interaction cross sections,  can either leave the detector or deposit their energy  late enough in time that the signal is not recorded.   Fluctuations in the number of nuclear breakups lead to fluctuations in the invisible energy, which impact the calorimeter signal.  Since neutral pions (which predominately decay promptly to two photons) produced in the nuclear breakups cannot initiate further breakups, and since their showers mostly produce signal via relativistic electrons/positrons, the amount of energy that does not produce a signal is anti-correlated with the fraction of electromagnetic (EM) objects in the shower.  
Showers with a large fraction of EM objects have an energy scale closer to that of electrons, those with a large number of hadronic interactions have a lower one, as illustrated in Fig.~\ref{fig:wigmans1} [left].
This variation in scale can be the dominant contribution to the energy resolution.

The dual readout technique uses two readouts, one sensitive to only energy deposits by the relativistic particles in the shower, the other sensitive to  deposits by all charged particles, including the heavy ion fragments and other slow particles produced in the nuclear breakups.  
Reference~\cite{Lee:2017xss} is a useful review with more details on the dual readout technique.

Dual readout is most often implemented in calorimeters that use optical media to produce signals.   Cherenkov radiation    is produced only by relativistic, charged shower particles, whose $\beta$ (speed divided by the speed of light)  is larger than the inverse of the index of refraction of the material. Scintillation light  is produced by all shower charged particles traversing scintillating media.  As discussed in detail in Ref.\cite{Lee:2017xss}, 
the scintillation signal $S$ and Cherenkov signal $C$ are related to  the true energy $E_{true}$ and each can be used as an energy estimator, after calibration.
If the calorimeter is calibrated using EM-objects such as electrons, the energy scale for hadrons is related  to
the fraction $f$ of the shower's energy deposits that  result from the shower products of EM objects   by 
\begin{equation}
S=E_{true} (f+(1-f)h_S),
\label{eqn:eq1}
\end{equation}

\begin{equation}
    C=E_{true} (f+(1-f)h_C).
\label{eqn:eq2}    
\end{equation}
where 
$h_C$ is the average response to the non-EM part of the shower for the $C$ measurement and $h_S$ is the same for the $S$ measurement.
In general, $h_C$ is less than $h_S$.

An improved energy estimator, that removes the dependence of the energy scale on $f$, can be obtained from Eqs.~\ref{eqn:eq1} and \ref{eqn:eq2}
The resulting dual-readout corrected energy estimate $D$ is
\begin{equation}
    D=\frac{(1-h_C)S-(1-h_S)C}{h_S-h_C}.
\end{equation}\label{eqn:dualE}
This equation can be rewritten as
\begin{equation}
    D=\frac{S-\chi C}{1-\chi},
\end{equation}\label{eqn:dualE2}
where
\begin{equation}
    \cot \theta \equiv \chi = \frac{1-h_S}{1-h_C}.
\end{equation}
The value of $\cot \theta$ is given by the slope of $C$ versus $S$.
As discussed in Ref.~\cite{Lee:2017xss},
this correction can be understood by considering  Fig.~\ref{fig:wigmans1} [right], and is equivalent to
moving the $S$-$C$ measurements along the line describing their correlation, to the electron energy scale.

Recently, results~\cite{chekanov2023geant4simulationssamplinghomogeneous} were presented that show that, instead of improving the resolution, the dual readout correction can make it worse.  As these results are at first counter intuitive, this paper clarifies under what conditions this is expected.  We derive a simple formula based on Eq.~\ref{eqn:dualE} that predicts the dual-readout-corrected energy from that estimated by the scintillator and Cherenkov signals alone.  As a byproduct, explicit formulae for $h_C$, $h_S$,  the scintillation resolution $\sigma_S$ and their relation to the escaping and binding energy losses are given in Sec.~\ref{sec:expand}.

\section{Resolution formula}
The $D$ resolution can be estimated from Eq.~\ref{eqn:dualE} using simple propagation of errors.  Since $S$ and $C$ are correlated via their dependence on $f$, the error propagation formula is

\begin{equation}
\sigma_D=\sqrt{\left(\frac{\partial E}{\partial S}\right)^2\sigma_S^2
+\left(\frac{\partial E}{\partial C}\right)^2\sigma_C^2
+2\frac{\partial E}{\partial S} \frac{\partial E}{\partial C} cov(S,C) 
},
\end{equation}
where $S$ and $C$ depend on $n$ random variables
$$
\Vec{x}=(x_1,x_2,...,x_n).
$$
Here $cov$ refers to the covariance,
and  $\sigma_S$ ($\sigma_C)$ is the measured resolution from the S (C) measurement.

The covariance  is given by
$$
cov(S,C)=\int_{-\inf}^{\inf} ... \int_{-\inf}^{\inf} S(\Vec{x})C(\Vec{x}) f(\Vec{x})dx_1dx_2...dx_n - <S><C>,
$$
where $f(\Vec{x})$ is the joint pdf for the $x_i$ and $<>$ represents the average value.

The calculation of the resolution is straightforward.

\begin{equation}
\frac{\partial D}{\partial S}=\frac{1-h_C}{h_S-h_C},
\end{equation}
 
\begin{equation} 
\frac{\partial D}{\partial C}=-\frac{1-h_S}{h_S-h_C}
\end{equation}

and the covariance is:

\begin{multline}
    cov(S,C) = \\
\int_{-\inf}^{\inf}  \int_{-\inf}^{\inf} \int_{-\inf}^{\inf}
((x_1+(1-x_1)h_S)+x_2)
((x_1+(1-x_1)h_C)+x_3)
g_1 g_2 g_3
dx_1 dx_2 dx_3 \\
- (<f>+(1-<f>)h_S)  (<f>+(1-<f>)h_C),\\
\label{eqn:eq9}  
\end{multline}
where $g_1$ is a Gaussian fluctuating $x_1$ with a mean of $<f>$ 
and an rms of $\sigma_f$
and which models fluctuations in the EM object portion of the shower,
$g_2$ is a Gaussian fluctuating  $x_2$ with a mean of zero and an rms of $n_S$ representing noise (or other non-correlated) resolution terms in the $S$ measurement,
$g_3$ is a Gaussian fluctuating  $x_3$ with a mean of zero and an rms of $n_C$ for the $C$ measurement.
As shown in~\ref{app:app1}, 
the result of the integration is
\begin{equation}
cov(S,C)=(1-h_S)(1-h_C)\sigma_f^2.
\label{eqn:cov}
\end{equation}

Putting it all together,
\begin{equation}
    \sigma_{D} = \frac{1}{h_S-h_C}\sqrt{(1-h_C)^2\sigma_s^2 +  (1-h_S)^2 \sigma_c^2
-2 (1-h_S)^2 (1-h_C)^2 \sigma_f^2}.
\label{eqn:maineqn}
\end{equation}

In terms of $\chi$, this can also be written
\begin{equation}
    \sigma_{D} = \sqrt{
    \frac{\sigma_s^2}{(1-\chi)^2}+\frac{\chi^2 \sigma_c^2}{(1-\chi)^2}
    -2 \frac{(h_S-h_C)^2 \chi^2 \sigma_f^2}{(1-\chi)^4}
    }.
    \label{eqn:resultchi}
\end{equation}

This formula works best if the $f$ is somewhat Gaussian.  The validity of this assumption will be discussed in Sec.~\ref{sec:calplots}.
It also works best if $f_{rms}$ is not too large, as we will discuss in Sec.~\ref{sec:toy}.
The formula  still gives insight even when these conditions are not satisfied.

The formula shows the conditions that are required for the dual readout correction to improve the resolution.
The resolution has two positive terms, one related to the $S$ resolution, the other to the $C$.  Both of these have contributions from $f$ variation, and in addition each has contributions from their own noise terms.  The $f$ variation is canceled by the second term.  However, if the noise contributions are larger than the contribution from the $f$ resolution, the dual readout energy estimate can have worse resolution than the individual measurements.  
Typically, the $S$ measurement is better as scintillation photon production and collection are more efficient. 
Looking at Eqn.~\ref{eqn:resultchi}, and remembering that, since the $C$ measurement is smaller than the $S$ measurement and thus $0<\chi<1$, the first term is larger than $\sigma_s$ by a factor of $1/(1-\chi)$.
The sum of the second and third terms will be positive, and make the $\sigma_D$ even worse compared 
to $\sigma_S$, unless
\begin{equation}
    \sigma_f>\frac{\sigma_C}{\sqrt{2}(1-h_C)}.
\end{equation}
The condition on $f$ to strictly require $\sigma_D>\sigma_S$ is
\begin{equation}
    \sigma^2_f>\frac{(2-\chi)\sigma^2_S}{2(1-h_c)(1-h_S)}+\frac{\sigma^2_C}{2(1-h_C)^2}.
\end{equation}

\section{Tests with a toy simulation}\label{sec:toy}

\begin{figure}[hbtp]
\centering
\resizebox{0.48\textwidth}{!}%
{\includegraphics{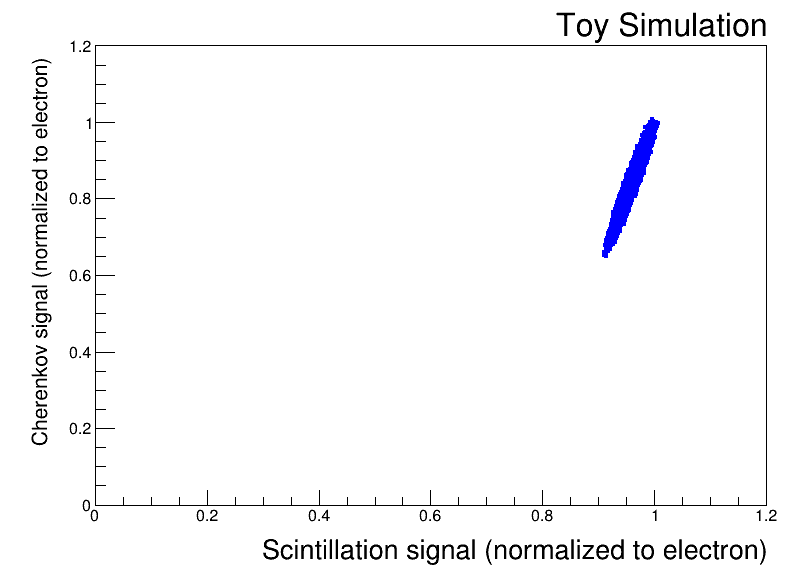}}
\resizebox{0.48\textwidth}{!}
{\includegraphics{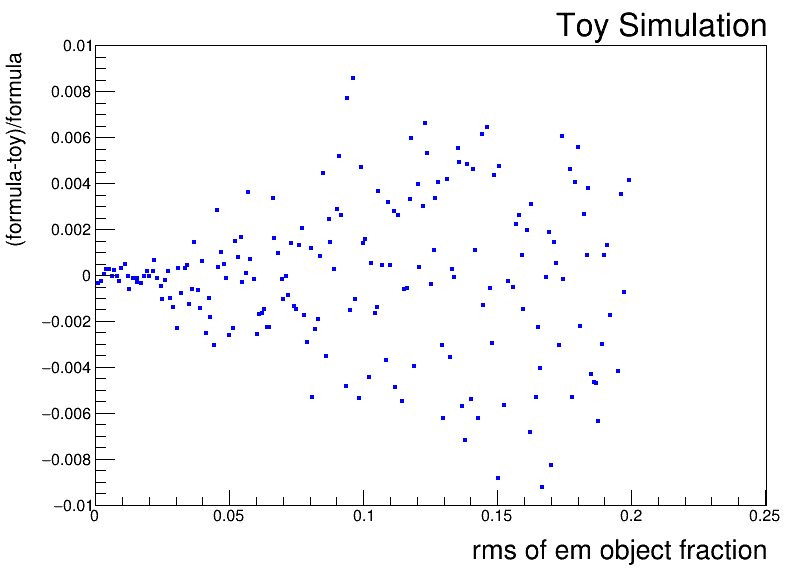}}
\caption{[left] Cherenkov versus scintillation signals from the toy MC using the parameters values defined in Sec.~\ref{sec:toy}  
[right] The percent difference between the dual-readout-corrected resolution measured from toy MC simulation ($\sigma_D$) and the computed resolution from Eq.~\ref{eqn:maineqn} as a function of the rms of the EM object fraction $f$. }
\label{fig:toy}
\end{figure}

We used a toy Monte Carlo simulation~\cite{toymc} to test Eq.~\ref{eqn:maineqn}. The toy models $f$ as a Gaussian.
The $S$ and $C$ signals are calculated from $f$ using Eqs.~\ref{eqn:eq1} and \ref{eqn:eq2}.  The $S$ and $C$ signals are then smeared using separate Gaussians to simulate sources of resolution uncorrelated between the two (photo statistics, sampling fraction fluctuations, {\it etc.}).
 Each toy generates 1000000 showers.
Fig.~\ref{fig:toy} [left] shows  $S$ versus $C$ for $h_S=0.9$, $h_C=0.6$,  $<f>=0.6$,   $\sigma_f=0.2$,
 and with a 0.3\% uncorrelated smearing on each of $S$ and $C$.
Fig.~\ref{fig:toy} [right] shows the fractional difference between the true and predicted dual-corrected resolution as a function of $f$.
The prediction agrees well as long as $\sigma_f$ is not too large.

\begin{figure}[hbtp]
\centering
\resizebox{0.5\textwidth}{!}
{\includegraphics{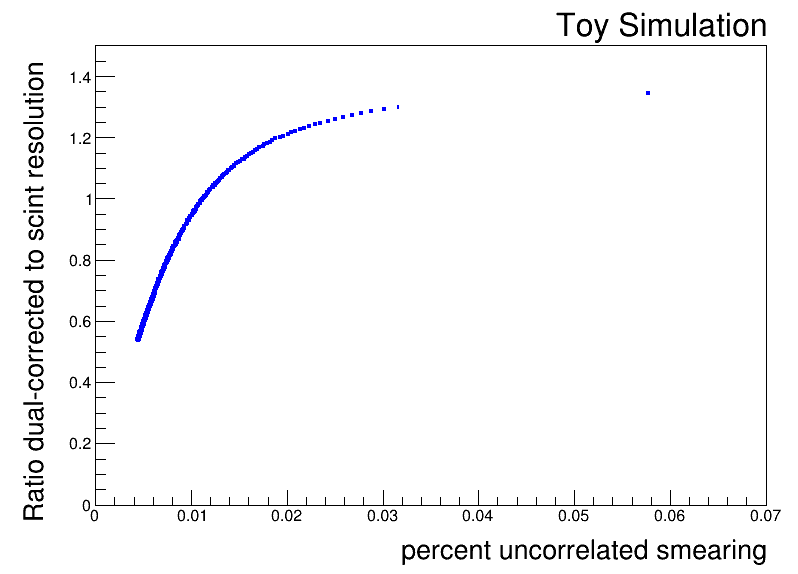}}
\caption{
Energy resolution from the dual-readout estimate divided by that from the scintillation measurement for $h_S=0.9$, $h_C=0.6$, $<f>=0.6$, and $\sigma_f=0.2$ when the uncorrelated smearing on $S$ and $C$ are varied.
}\label{fig:neg}
\end{figure}

Fig~\ref{fig:neg} shows the ratio of the resolutions of the $D$ and $S$ measurements when the uncorrelated smearing on the $S$ and $C$ is varied for $h_S=0.9$, $h_C=0.6$, $<f>=0.6$, and $\sigma_f=0.2$.  In this case, the the $D$ resolution is better than $S$ alone when the noise term is less than about 1\%.

\section{Comparison with full simulation results}

\subsection{Simulation geometries and conventions}\label{sec:geometries}

\begin{figure}[hbtp]
\centering
\resizebox{0.80\textwidth}{!}{\includegraphics{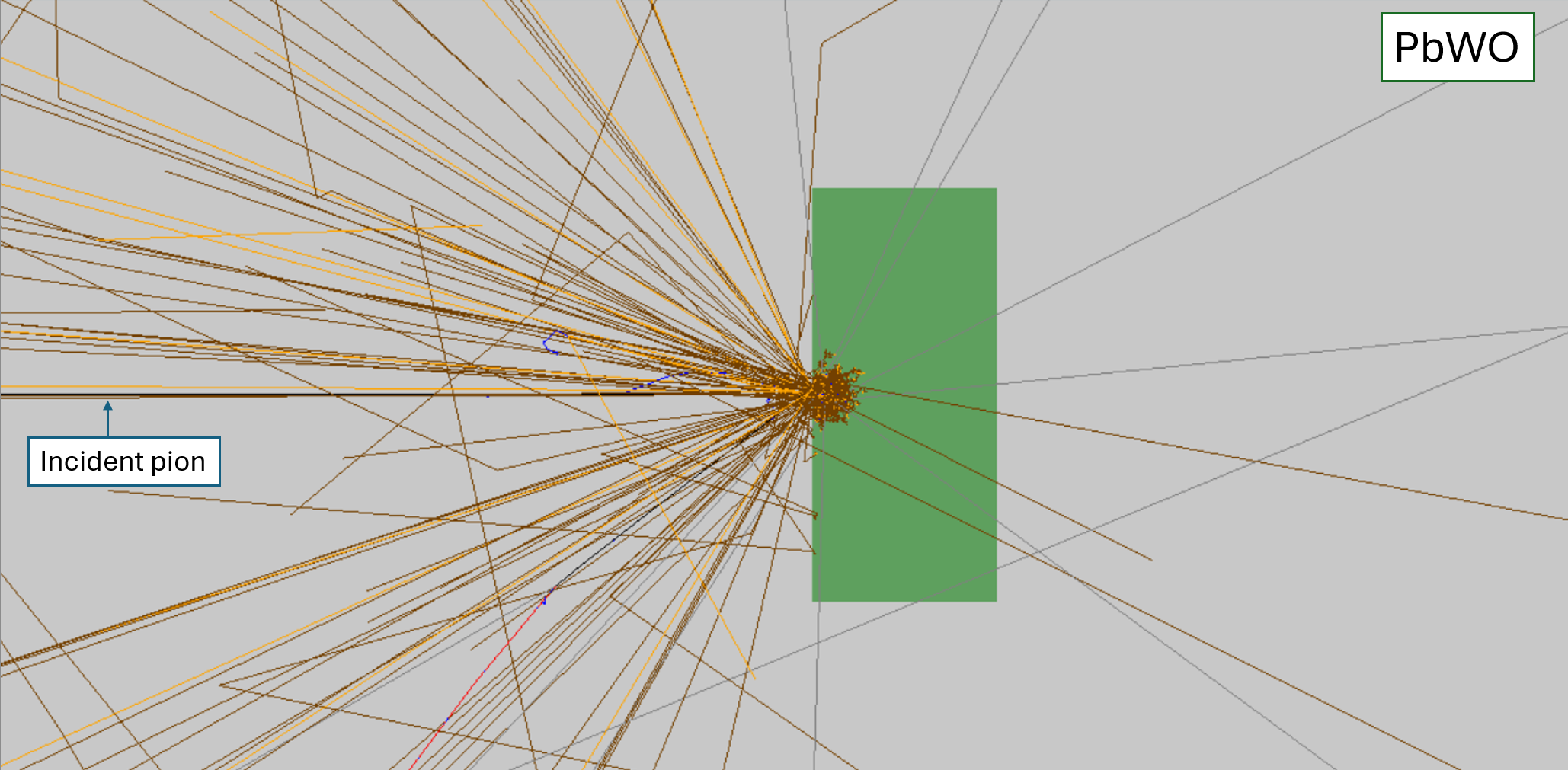}}
\caption{ Shower from a 20 GeV charged pion for
crystal calorimeter PbWO.
The colors for particles are: electrons/positrons blue, charged pions black, muons red, neutrons brown, and photons orange.
The colors for the detector are: hall grey and PbWO4 green.
}\label{fig:theCals1}.  
\end{figure}

\begin{figure}[hbtp]
\centering
\resizebox{0.80\textwidth}{!}{\includegraphics{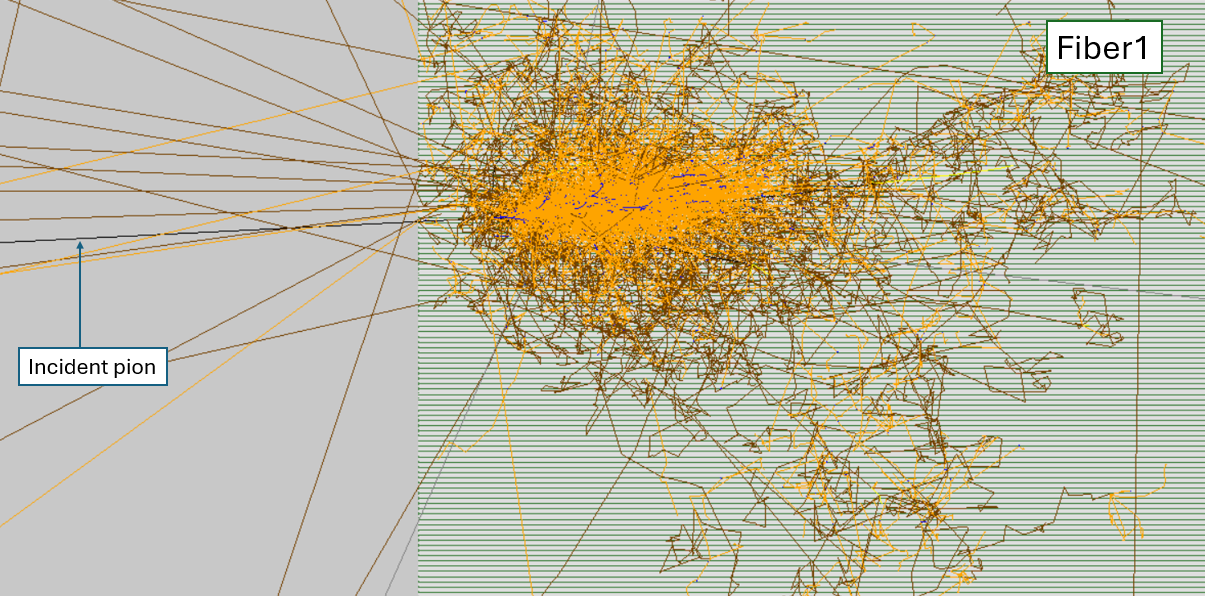}}
\resizebox{0.80\textwidth}{!}{\includegraphics{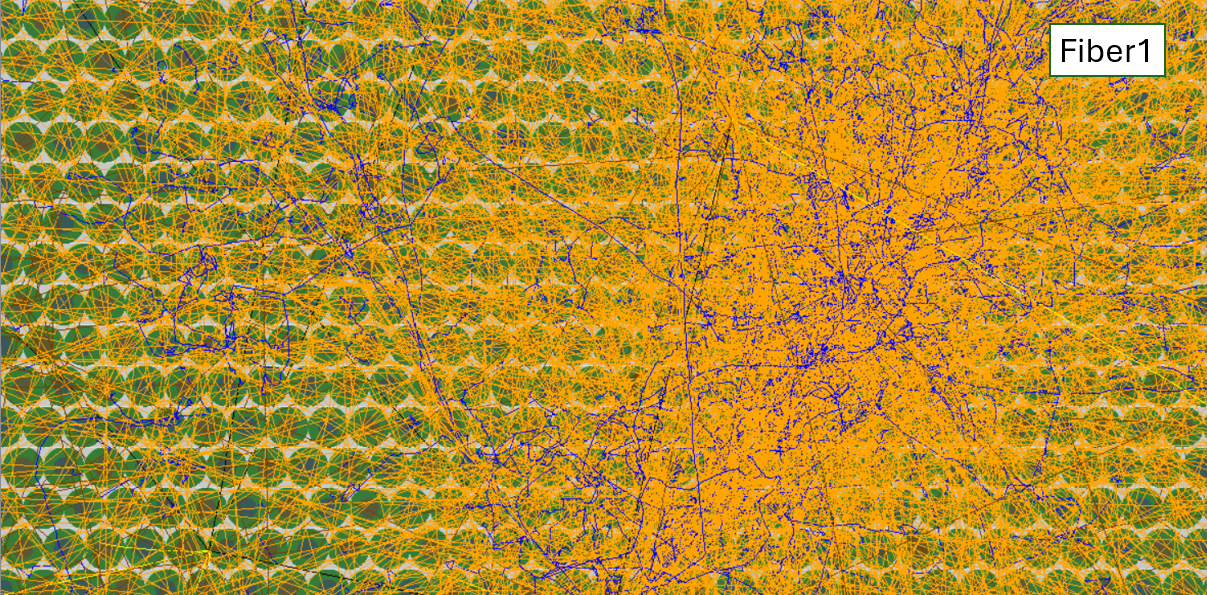}}

\caption{ Shower from a 20 GeV charged pion for
 Fiber calorimeter Fiber1.
The colors for particles are: electrons/positrons blue, charged pions black, muons red, neutrons brown, and photons orange.
The colors for the detector are: hall grey and PbWO4 green.
The top image shows the y-z view, where z is along the calorimeter depth.  The bottom shows a close-up of the x-y view.
}\label{fig:theCals3}.  
\end{figure}

\begin{figure}[hbtp]
\centering
\resizebox{0.80\textwidth}{!}{\includegraphics{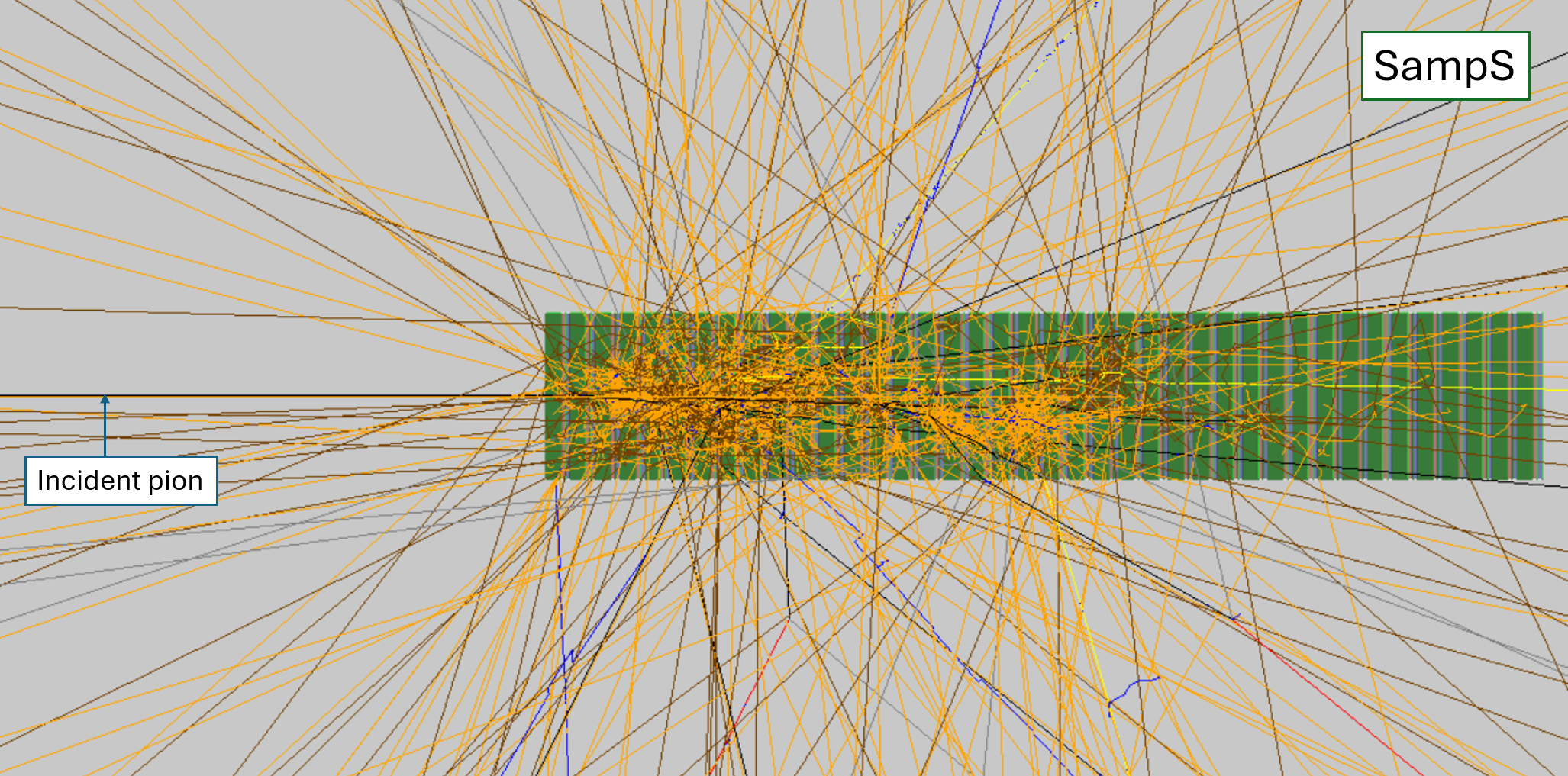}}
\caption{ Shower from a 20 GeV charged pion for
Tile calorimeter SampS.
The colors for particles are: electrons/positrons blue, charged pions black, muons red, neutrons brown, and photons orange.
The colors for the detector are: hall grey and PbWO4 green.
}\label{fig:theCals2}.  
\end{figure}

To further test and  expand on the equation, we simulate up to two variations of three different types of calorimeters using GEANT4~\cite{geant4} wrapped in DD4HEP~\cite{dd4hep}.  We simulate an unaffordably-large block of scintillating crystal, referred to as calorimeter PbWO.  We also simulatate two  traditional tile-based sampling calorimeter such as the CMS hadron calorimeter~\cite{HCalBarrel} except with both scintillating and Cherenkov layers, calorimeters SampL and SampS.  Calorimeter SampS is narrow and short, with significant escaping energy.  Finally, we simulate 
 two spaghetti-type fiber calorimeter similar to that proposed by the IDEA collaboration~\cite{Bedeschi:2021nln}, with separate scintillating and Cherenkov fibers inside individual absorber tubes, calorimeters Fiber1 and Fiber2.  Fiber2 has larger diameter fibers than Fiber1.  
Tables~\ref{tab:parameters} and \ref{tab:parameters2}  give the labels and  geometric dimensions for all variations.
Figures~\ref{fig:theCals1}, \ref{fig:theCals2}, and \ref{fig:theCals3} show event displays for the PbWO, SampL, and Fiber1 calorimeters respectively.
Calorimeter SampS was also simulated for Ref.~\cite{chekanov2023geant4simulationssamplinghomogeneous}.
Calorimeter Fiber1 is similar to that simulated in Ref.~\cite{Lucchini:2020bac}.

Particles enter the calorimeter perpendicular to its face for the crystal and tile calorimeters. For fiber calorimeters, to avoid a particle aligned with a fiber,  the particles enter the calorimeter with an angle of about 3$^o$~with respect to the perpendicular.
 Electrons and negatively charged pions with energies of 20 GeV were simulated for each calorimeter. 
 For each particle type, 2000 showers were simulated.

 Energy deposits in the active media were used as the signal for most of the results.
 For Cherenkov signals,
we sum energy deposits by relativistic particles, defined  as having $\beta>1/1.5$, as this is a typical index of refraction for media producing Cherenkov signals.
At a wavelength of 410\,nm, the indices of refraction for ${\rm PbWO_4}$, quartz, and polystyrene are
2.4, 1.4, and 1.6 respectively.
This method of signal estimation is equivalent to large photon collection efficiency, and so is equivalent to neglecting the contribution of photostatistics.  It does also neglect  suppression of photons due to Birke's Law~\cite{birks} for slow, highly-ionizing particles. Ion fragments from nuclear breakups are the most common particle of this type.
As a check, results obtained by counting created photons (without propagation to photodetectors) are included for the resolution plots
shown in Fig.~\ref{fig:threeres}.
Cherenkov photons are created for wavelengths between 200 and 999\,nm.  The resulting response distributions are identical for the $S$ and $D$ resolutions.  There are small differences for the Cherenkov signals due to our rough definition of relativistic that uses a single $\beta$ selection requirement for the all Cherenkov materials.

The $S$ and $C$ signals are calibrated by normalizing to the most probable signal for electrons.

\begin{table}[!htp]
\centering
\caption{Labels and parameters of simulated calorimeters. 
The columns give the label, the calorimeter type, the scintillator material, the scintillator thickness, the Cherenkov material, the Cherenkov thickness, the absorber material, and the absorber thickness.
The label "thick" is either the thickess or outer diameter. The label ``n.a." means not applicable.   }
\begin{tabular}{c c c c c c c c}
\hline
Label & Type    & Scint. & Scint. & Cher. & Cher. & Abs. &Abs \\
    &        &  mat. & thick  & mat.  & thick & mat.  &thick \\
      &  & & cm & & cm& & cm \\      
         \hline
PbWO & crystal & ${\rm PbWO_4}$ & n.a. & n.a.&n.a. & n.a.&n.a. \\
Fiber1 & fiber & Polystyrene & 0.05 & Quartz & 0.05 & Brass & 0.1 \\
Fiber2 & fiber & Polystyrene & 0.05 & Quartz & 0.05 & Brass & 0.075 \\
SampL & tile & Polystyrene & 0.5 & Quartz & 0.5 &Iron & 1.8 \\
SampS & tile & Polystyrene & 0.5 & Polystyrene & 0.5 &Iron & 1.8 \\
\hline
\end{tabular}
\label{tab:parameters}
\end{table}

\begin{table}[!htp]
\centering
\caption{Labels and parameters of simulated calorimeters.  The columns are the calorimeter label, its depth, and its width}
\begin{tabular}{c c c }
\hline
Label & Depth & Width  \\
      &  cm & cm \\
         \hline
PbWO & 405 & 900 \\
Fiber1 & 210 &  50.1 \\
Fiber2 & 210 &  54.075 \\
SampL & 224 &  420 \\
SampS & 112 &  20 \\
\hline
\end{tabular}
\label{tab:parameters2}
\end{table}

\subsection{Expansion of formula}\label{sec:expand}
When dividing a shower into components, it is natural to divide it in terms of contributions from the relativistic and non-relativistic particles in the shower, since only the relativistic fraction produces the Cherenkov signal.  However, in terms of calibration and the sources of lower response, it is more natural to write it in terms of the EM object part and the rest (non-EM).   To give further insight into dual readout calorimeter resolutions, we make use of both quantities.

\begin{figure}[hbtp]
\centering
\resizebox{0.60\textwidth}{!}{\includegraphics{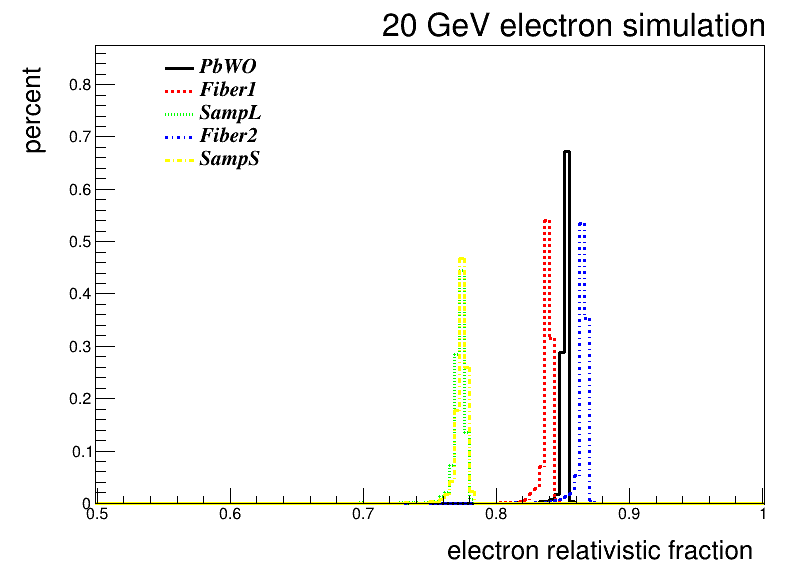}}
\caption{ Fraction of deposited ionizing energy in 20 GeV electron showers deposited by particles with $\beta>1/1.5$ for the various calorimeters.}\label{fig:relEe} 
\end{figure}

Consider the shower of an EM object such as an electron or photon.  Even in such a shower, not all  particles are relativistic. The relativistic fraction of the shower, and its relation to the energy deposits from relativistic and from all shower particles, is defined as follows:
\begin{equation}
    f_{R, \gamma}= \frac{E_{R, \gamma}}{E_B}
\end{equation}
where  $E_B$ is the true particle energy and
 $E_{R, \gamma}$ is the ionizing energy deposition by relativistic shower particles in the EM shower.  Here we assume the calorimeter is large enough that the energy escaping the calorimeter for incident electrons and photons is negligible.  This assumption will not be made for hadronic showers.
The relativistic fraction for electron showers is shown in Fig.~\ref{fig:relEe}.
The fraction of ionizing energy deposited by relativistic particles is lower for calorimeters SampL and SampS, which are based on iron, then for Fiber1 and Fiber2, which are based on brass, or PbWO, which is based on lead-tungstate.  
Iron has the lowest atomic number, and thus at photon energies just below the pair-production threshold,
has a larger contribution from Compton scattering compared to the photoelectric effect.

Now consider a pion shower. The fraction of energy of deposits by relativistic particles in the shower  $f_{R,\pi}$ is defined by analogy. 
\begin{equation}
    f_{R, \pi}=\frac{E_{R,\pi}}{E_B},
\end{equation}
where $E_{R,\pi}$ is the energy deposited by  relativistic particles in a pion  shower.
If we assume that the energy depositions by relativistic particles are all due to particles from EM objects
(the simulation results will test the validity of this assumption),
the energy in the shower from EM objects is then
\begin{equation}
    E_{\gamma,\pi}=E_{R,\pi}/f_{R,\gamma}.
\end{equation}

We define $f_{\gamma,\pi}$ as
\begin{equation}
    f_{\gamma,\pi}=\frac{f_{R,\pi}}{f_{R,\gamma}}.
\end{equation}
This is thus a conversion factor from the beam energy, then to  the relativistic fraction (numerator), and to the EM object fraction (denominator).

Consider now the $S$ signal (the $C$ follows by analogy). 
Dividing the shower into an EM and non-EM portion,
we postulate an ansantz:
\begin{equation}
    S=g_1 f_{\gamma,\pi} E_B + g_2[(1-f_{R,\pi})E_B -  (1-f_{R,\gamma})f_{\gamma,\pi} E_B](1-f_{NC}-f_{ES})
    \label{eqn:ansatz}
\end{equation}
where the first term is the EM-object portion of the shower and the second term is the rest of the shower.  Inside the square brackets, the second term removes the non-relativistic portion of the EM portion of the shower to avoid double counting with the first term.  The $g_1$ and $g_2$ terms are the ratios of the sampling fractions for the EM and non-EM parts of the shower to that for electrons.  For a homogeneous calorimeter, such as  PbWO $g_1$ and $g_2$ are one.  The derivation for these sampling calorimeters is given later.
It also has a term related to the fraction of energy loss which does not produce signals, due to binding energies for nuclear breakups ($f_{NC}$), and due to energies that escape from the calorimeter ($f_{ES}$). 
We postulate these terms have no effect on the EM part of the shower.  We will show this is true as long as the calorimeter is not very small.
Technically, $f_{NC}$ is calculated as
\begin{equation}
    f_{NC}=\frac{E_B-E_{ES}-E_{I}}{E_B}
\end{equation}
where $E_{ES}$ is the energy escaping the calorimeter and $E_{I}$ is the sum of all ionizing energy deposits.

From examining this equation, a useful quantity is 
\begin{equation}
    E_{{\rm non}EM}= [(1-f_{R,\pi})E_B - (1-f_{R,\gamma})f_{\gamma,\pi}E_B]
    \label{eqn:nonem}
\end{equation}
which is our definition of the non-EM part of the shower.

Assuming $g_1$ is one, $S$ can be rewritten as
\begin{equation}
    S=f_{\gamma,\pi} E_B + g_2(1-f_{\gamma,\pi}) (1-f_{NC}-f_{ES})E_B
    \label{eqn:ansatz2}
\end{equation}
comparing to Eq.\ref{eqn:eq1}, we can identify 
\begin{equation}
    f \equiv f_{\gamma,\pi} = \frac{f_{R,\pi}}{f_{R,\gamma}}
\end{equation}
and
\begin{equation}
h_S \equiv g_2 (1-<f_{NC}>-<f_{ES}>)
\label{eqn:hs2}
\end{equation}

Analogously, remembering the detector is calibrated to unit response for electrons/photons,
\begin{equation}
    C=\frac{f_{R,\pi}}{f_{R,\gamma}}E_B
    \label{eqn:cpred}
\end{equation}
and $h_C$ is zero.  Note that $h_C$ is zero only using the EM object fraction of the shower properly defined, not when using the relativistic fraction of the shower.

The resolution for $S$ can be then obtained via propagation of errors.  It is the quadratic sum of two components, one related to the variation of $f$ and another that does not correlate the EM object and non-EM orthogonal portions of the shower.
The part of the resolution driven by $f$ is
\begin{equation}
    \sigma_{S,corr}=(<f_{NC}>+<f_{ES}>)\sigma_f.
    \label{eqn:res1}
\end{equation}
The non-correlated part is given by
\begin{equation}
    \sigma_{S,non-corr}=(1-<f>)\sqrt{\sigma^2_{f_{NC}}+\sigma^2_{f_{ES}} + (\frac{\sigma_g}{g})^2 },
    \label{eqn:res2}
\end{equation}
where $\sigma_g$ is the rms of the sampling fraction distribution.
The resolution is thus
\begin{equation}
    \sigma_S=\sqrt{\sigma^2_{S,corr} + \sigma^2_{S,non-corr} }.
    \label{eqn:scintrespred}
\end{equation}
This formula is only valid if $g_2$, $f$, $f_{ES}$, and $f_{NC}$ are uncorrelated.
However, $g_2$ and $f_{NC}$ should in general be correlated.
Nevertheless, it is a good approximation of the resolution for most calorimeters (see Sec.~\ref{sec:results}).

\subsection{Calorimeter behaviors}\label{sec:calplots}
In this section, we look at the behavior of some of the relevant quantities in Eqs.~\ref{eqn:ansatz2} to \ref{eqn:res2}.  The behaviors will be very different, and the efficacy of the dual readout correction will vary, but the trend will be predicted well by the Eq.~\ref{eqn:maineqn}, as will be  shown in Sec.~\ref{sec:results}.

\begin{figure}[hbtp]
\centering
\resizebox{0.550\textwidth}{!}{\includegraphics{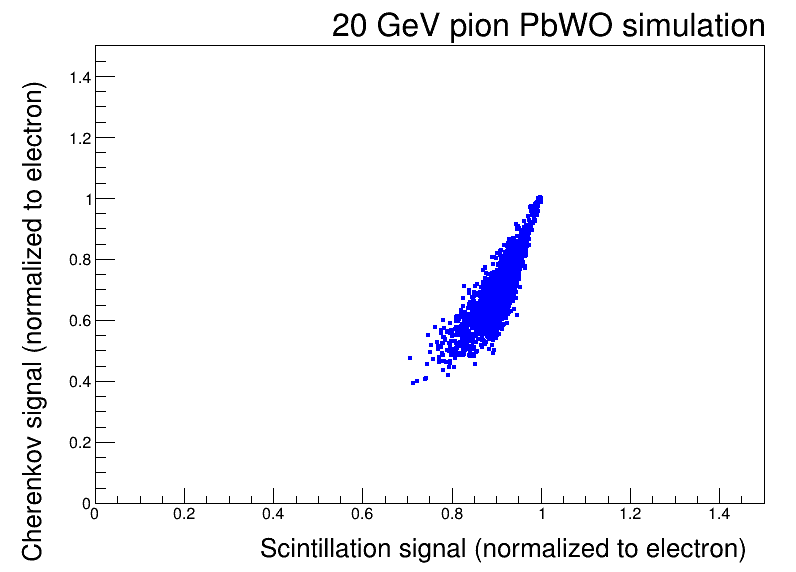}}
\resizebox{0.480\textwidth}{!}{\includegraphics{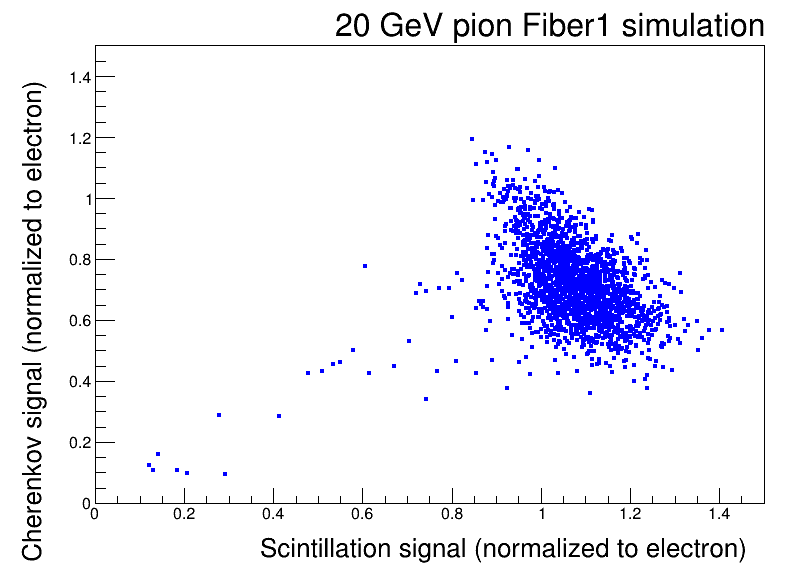}}
\resizebox{0.480\textwidth}{!}{\includegraphics{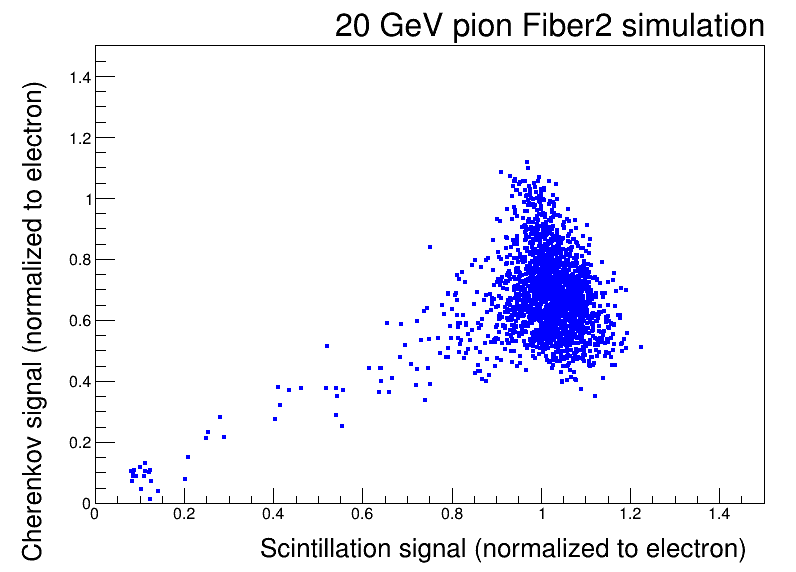}}
\resizebox{0.480\textwidth}{!}{\includegraphics{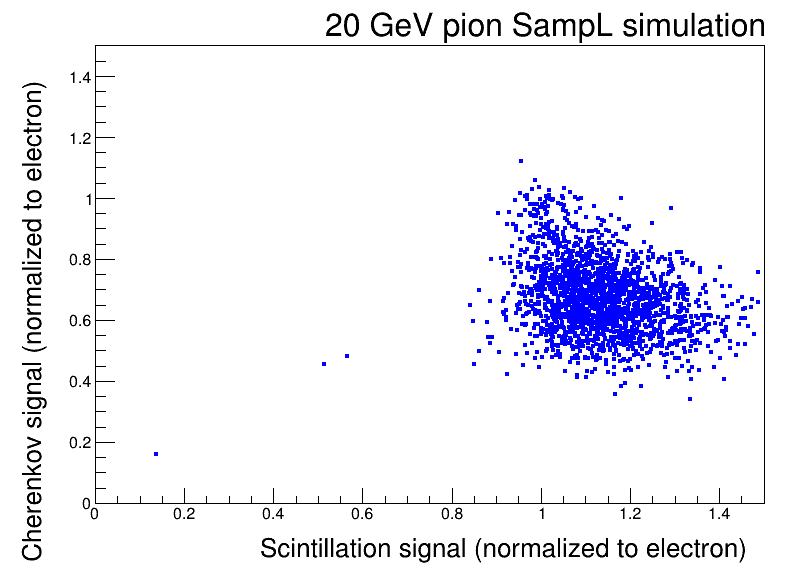}}
\resizebox{0.480\textwidth}{!}{\includegraphics{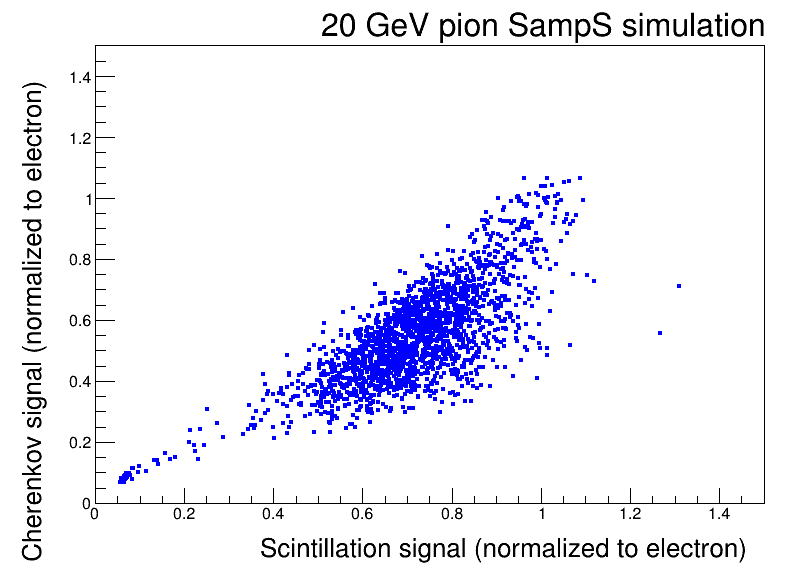}}

\caption{ 
Cherenkov versus scintillation signal for calorimeters
[top] PbWO,
[middle] Fiber1, Fiber2,
[bottom] SampL, SampS.
 }\label{fig:nsnc}.  
\end{figure}

\begin{figure}[hbtp]
\centering
\resizebox{0.60\textwidth}{!}{\includegraphics{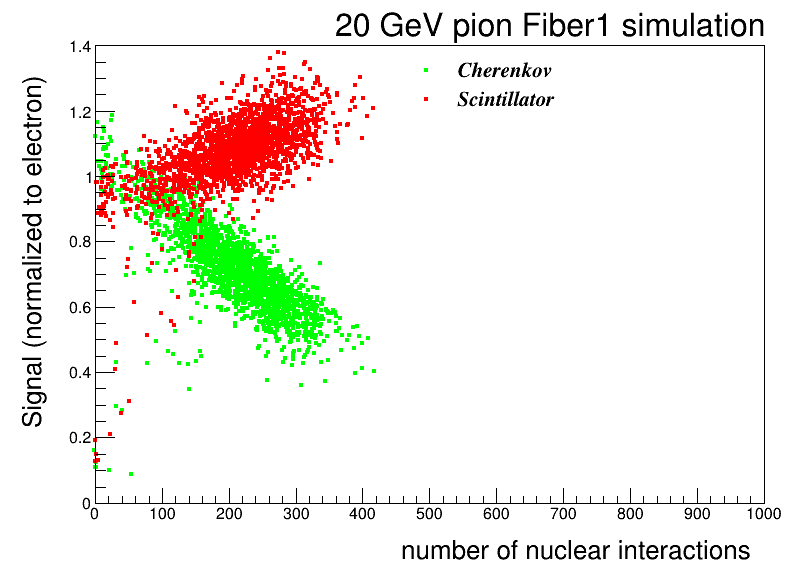}}
\caption{ 
Cherenkov (green) and scintillation (red) signal versus number of nuclear interactions for calorimeter Fiber1.
 }\label{fig:nsncnuci}.  
\end{figure}

\begin{figure}[hbtp]
\centering
\resizebox{0.480\textwidth}{!}{\includegraphics{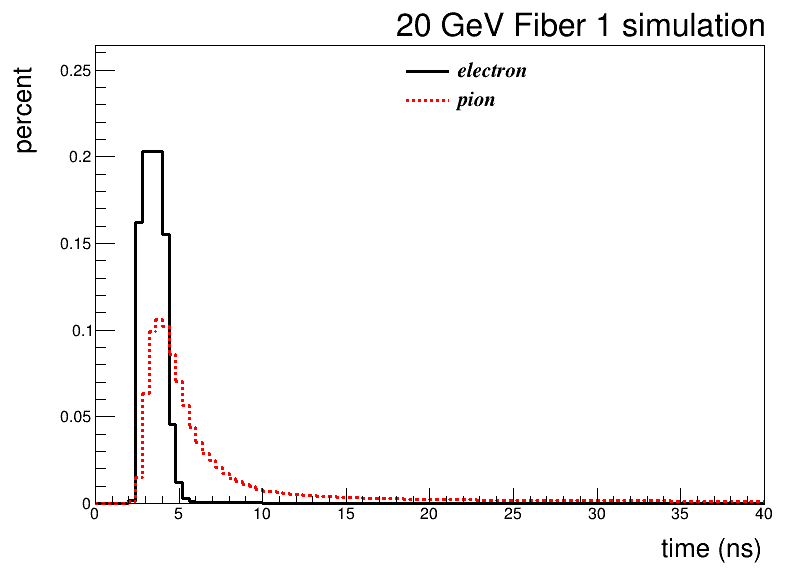}}
\resizebox{0.480\textwidth}{!}{\includegraphics{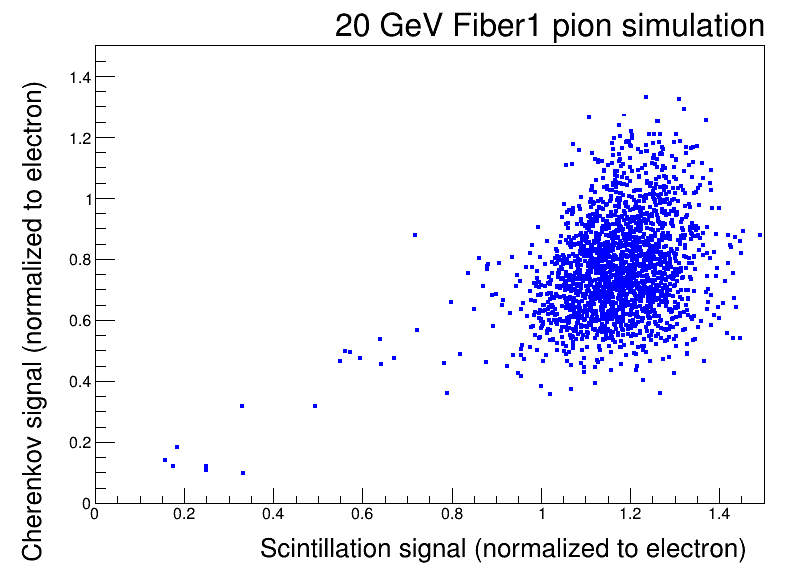}}
\caption{ For calorimeter Fiber 1,
 [left] the distribution of the times of energy deposits for electrons (black) and pions (red), and
 [right] Cherenkov signal versus scintillation signal for those signals with deposition time less than 10\,ns.
 }\label{fig:timingcut}.  
\end{figure}

The canonical scatter plot of $S$ versus $C$ for each calorimeter is shown in Fig.~\ref{fig:nsnc}.  While PbWO and SampS show the behavior most often pictured, with a positive slope for $C$ versus $S$, SampL, Fiber1, and Fiber 2 show a negative slope.  This occurs because both the tile and fiber calorimeters contain hydrogen-rich materials (polystyrene). Hydrogen is the only material in these calorimeters with a substantial neutron interaction cross section. As the number of nuclear interactions increases (and thus the energy lost to nuclear breakups increases), more neutrons are produced.
The effect of this on the signals can be seen in Fig.~\ref{fig:nsncnuci}, which shows $S$ and $C$ versus the number of nuclear breakups in the shower. 
 This simulation integrates the signal over all time, and includes even late time signals from the neutrons.  The hydrogen-rich material boosts the neutron signal, and thus the response for $S$ increases with larger number of nuclear breakups, while $C$ decreases.  This is illustrated in Fig.~\ref{fig:timingcut} for calorimeter Fiber1.  The left-hand plot shows the timing distribution of the energy deposits.  The right shows the scatter plot when including only signals with a time less than 10\,ns, which will remove mostly signals from neutrons.  The slope changes towards less negative.
 While SampS also contains hydrogen-rich sampling material, its slope is positive even before a timing cut; as can be seen in Fig.~\ref{fig:nonCons}, most of the non-EM energy (which includes the neutrons) escapes the calorimeter for this geometry.

\begin{figure}[hbtp]
\centering
\resizebox{0.480\textwidth}{!}{\includegraphics{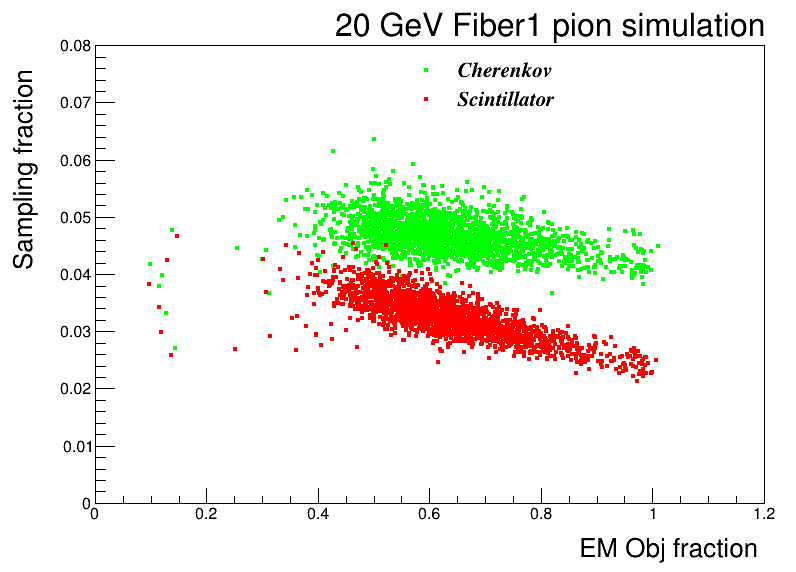}}
\resizebox{0.480\textwidth}{!}{\includegraphics{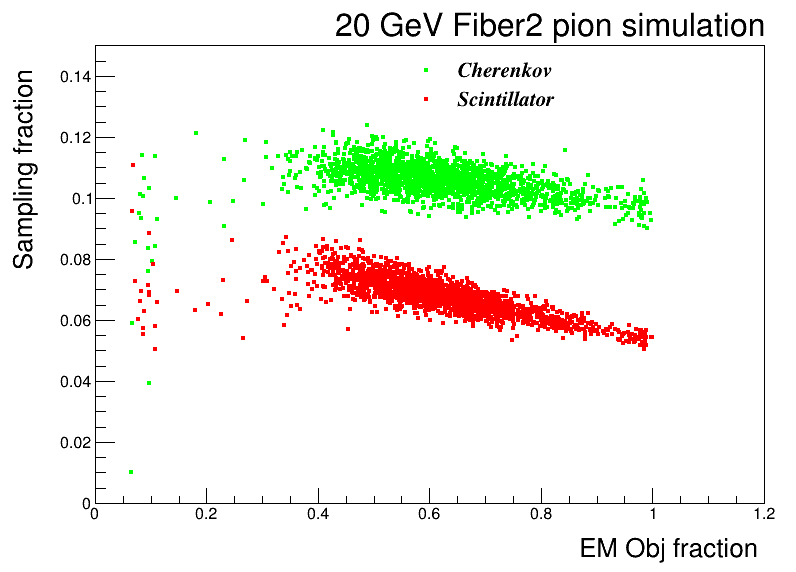}}
\resizebox{0.480\textwidth}{!}{\includegraphics{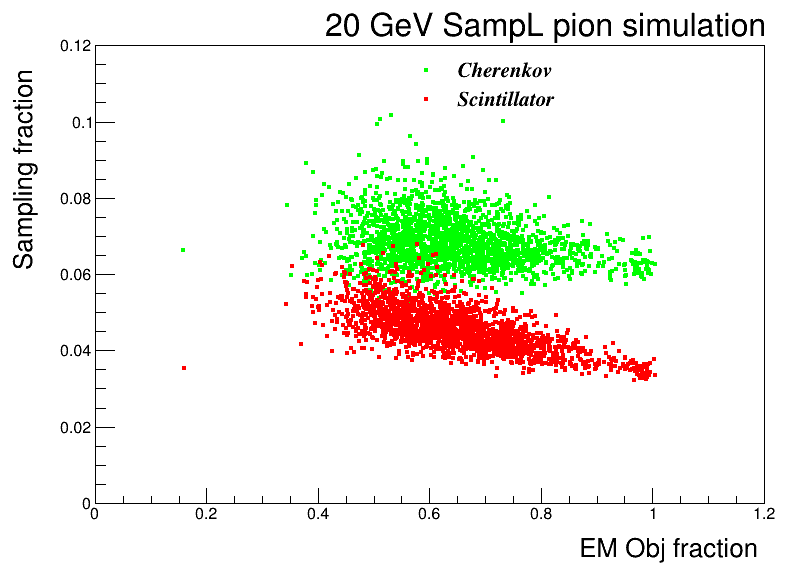}}
\resizebox{0.480\textwidth}{!}{\includegraphics{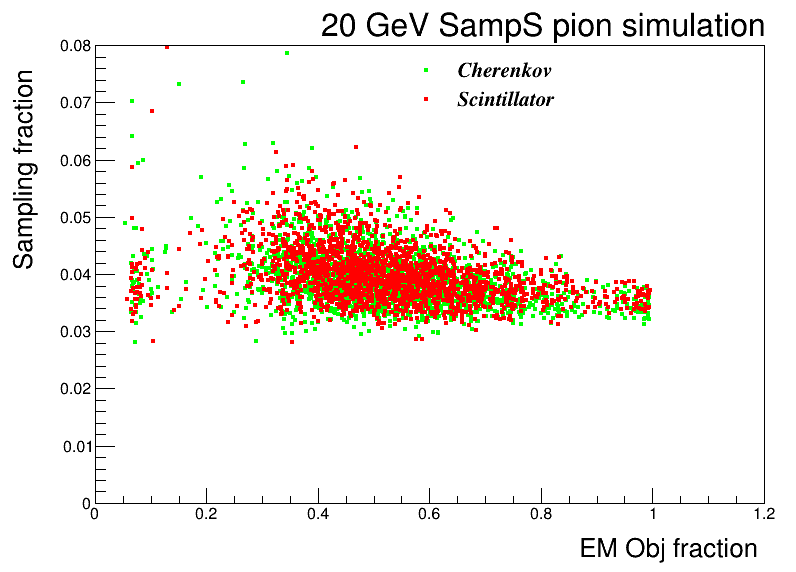}}
\caption{ 
Sampling fraction versus $f$ for the scintillating (red) and Cherenkov (green) active media for calorimeters
[top, left] Fiber1,
[top, right] Fiber2,
[bottom, left] SampL,
[bottom, right] SampS.
 }\label{fig:gvf1}.  
\end{figure}

The sampling fractions versus $f$ for the sampling calorimeters are shown in Fig.~\ref{fig:gvf1}.  
The values depend on whether  the active material contains hydrogen. 
For our calorimeters, Fiber1, Fiber2, and SampL have hydrogen in their scintillation but not their Cherenkov, media.  Because of this, the sampling fraction for $S$ has a stronger slope with the shower EM object fraction than for $C$.
The larger sampling fraction near an EM object fraction of one for $C$ is due to the heavier elements in quartz compared to polystyrene.
SampS has hydrogen in both media.   Due to its small size, however, most neutrons escape the calorimeter, resulting in a small slope with shower EM-object fraction.
Note that the sampling fraction for Fiber2 is about twice that of Fiber1.  SampS and SampL have intermediate sampling fractions.
If $g_1$ and $g_2$ are constant, with low dependence on $f$, the sampling fraction will have a linear dependence on $f$.  This is consistent with the observed distributions.
The values of $g_1$ and $g_2$ can be extracted by fitting to a line and setting $g_1$ to the result for $f=1$, $g_2$ for $f=1$, divided by the electron sampling fraction.  The results and comparison to the values for electrons is given in Tab.~\ref{tab:calibc}.  

\begin{figure}[hbtp]
\centering
\resizebox{0.55\textwidth}{!}{\includegraphics{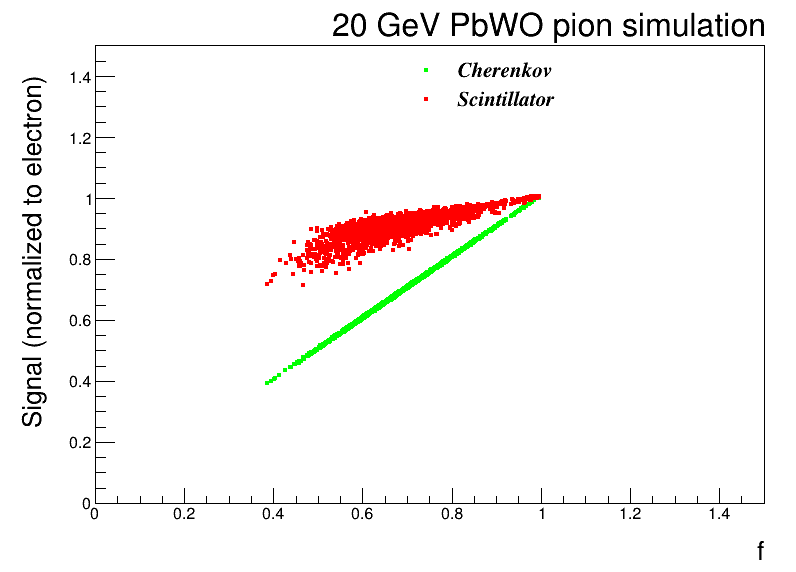}}
\resizebox{0.48\textwidth}{!}{\includegraphics{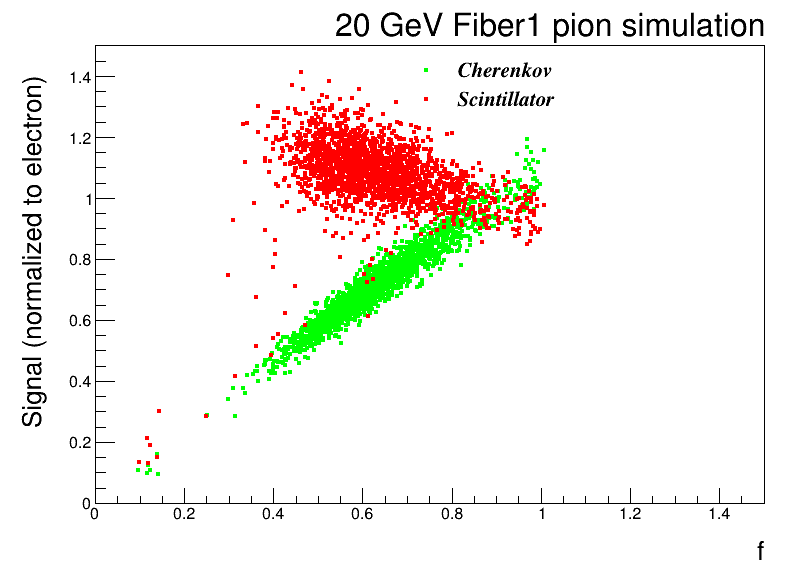}}
\resizebox{0.48\textwidth}{!}{\includegraphics{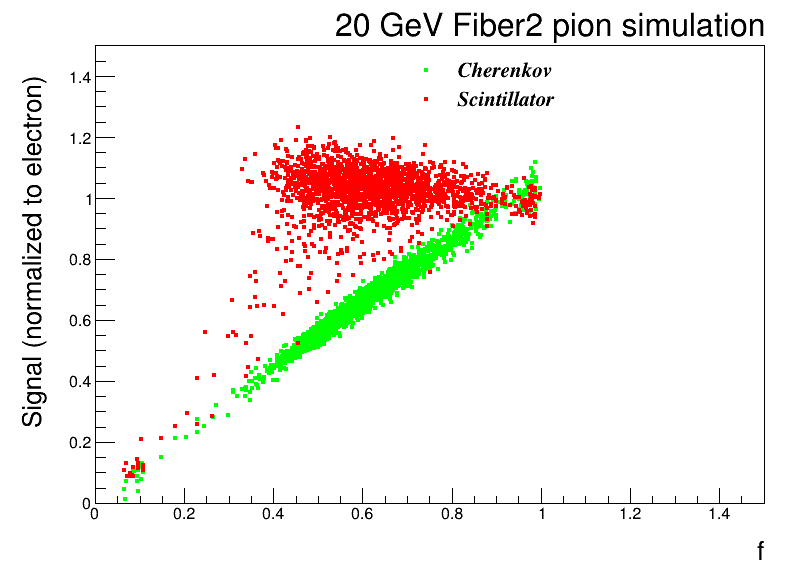}}
\resizebox{0.48\textwidth}{!}{\includegraphics{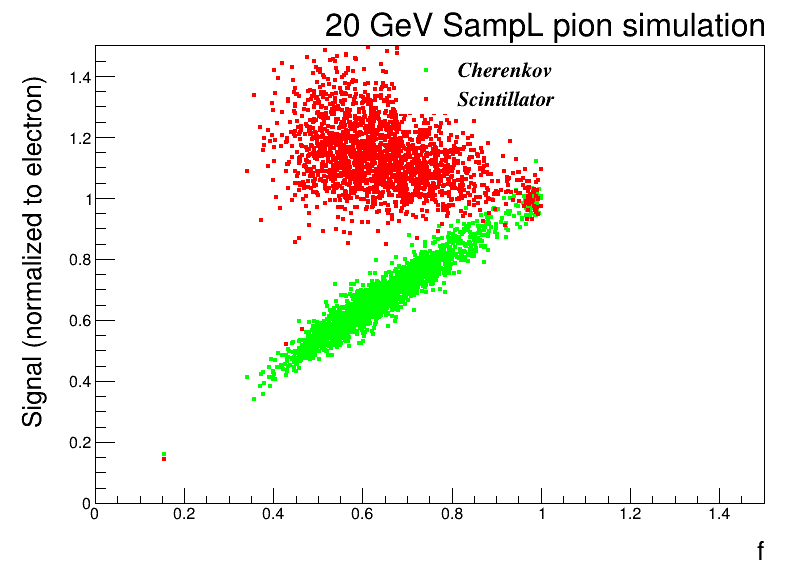}}
\resizebox{0.48\textwidth}{!}{\includegraphics{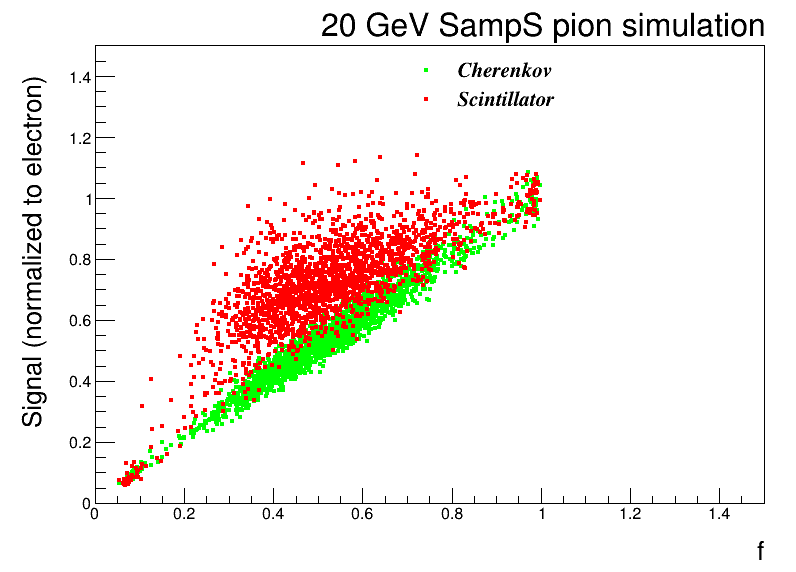}}
\caption{ 
Scintillation (red) and Cherenkov (green) signals versus the EM object fraction $f$ of the shower for calorimeters
[top] PbWO,
[middle] Fiber1, Fiber2,
[bottom] SampL, SampS.
 }\label{fig:calwoes2}.  
\end{figure}

For the dual-readout correction to work optimally, both $S$ and $C$ should be linear in $f$.
Figure~\ref{fig:calwoes2} shows the relationship.
The linearity assumption is reasonable.
Note that the narrowness of the $C$ distribution for homogeneous calorimeter PbWO (albeit with perfect scintillation/Cherenkov discrimination).  For the other calorimeters, the width is dominated by the sampling fraction.
A calorimeter is compensating if the gain from neutrons balances the loss from binding energy and escaping energy as $f$ varies.  The scintillation response versus $f$ for calorimeter Fiber2 is close to compensating, showing low variation with $f$.  For such a calorimeter, the dual-readout correction should not improve the resolution, as the $f$ variation is already removed.
The ansatz of Eq.~\ref{eqn:cpred} predicts $h_c=0$.
The offset is very small for calorimeter PbWO ($h_c$=0.006), SampL ($h_c$=0.08), and SampS ($h_c$=0.04). For calorimeters Fiber1 ($h_c$=0.21) and Fiber2 ($h_c$=0.14) the offset is larger.
An offset occurs when the second term in Eq.~\ref{eqn:nonem} does not completely remove all the non-relativistic energy of the EM objects or when the non-EM-object portion of the shower does produce relativistic particles.
For the fiber calorimeters, where the directionality of the shower has a strong influence on the sampling fraction, our calibration procedure of using electrons and pions with the same incident angle may be naive. 

\begin{figure}[hbtp]
\centering
\resizebox{0.550\textwidth}{!}{\includegraphics{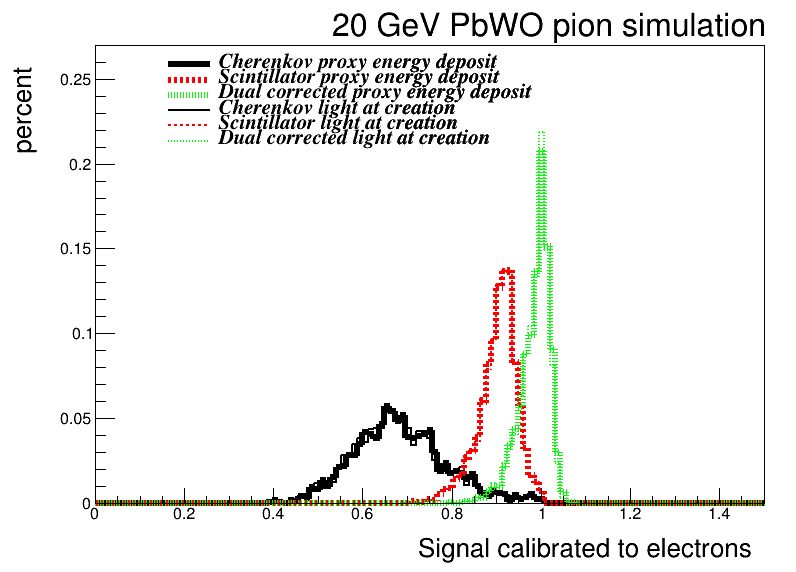}}
\resizebox{0.480\textwidth}{!}{\includegraphics{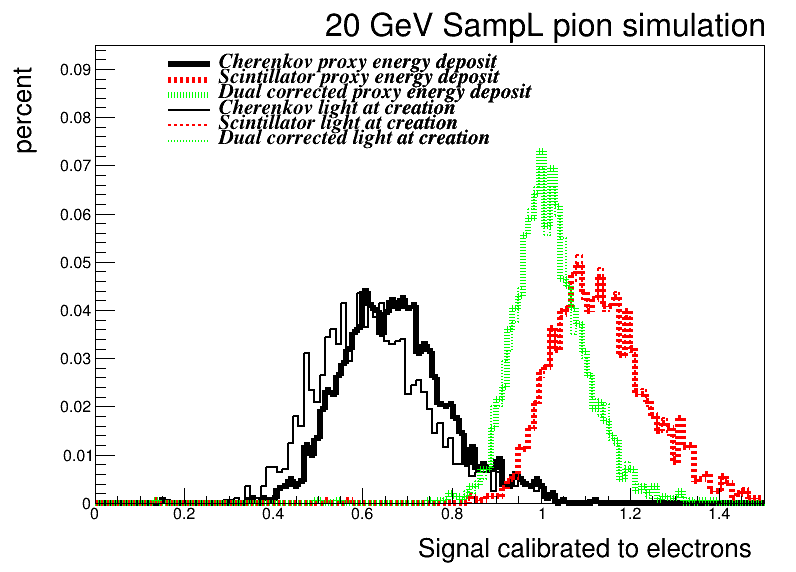}}
\resizebox{0.480\textwidth}{!}{\includegraphics{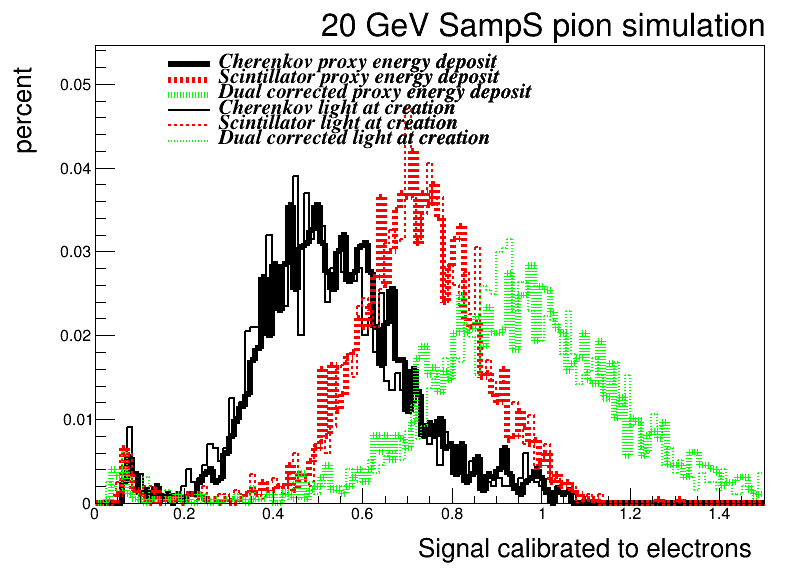}}
\resizebox{0.480\textwidth}{!}{\includegraphics{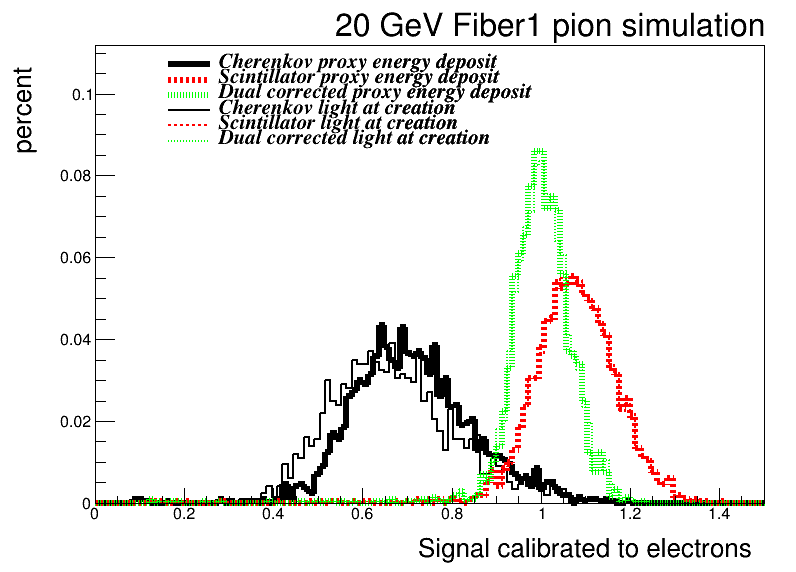}}
\resizebox{0.480\textwidth}{!}{\includegraphics{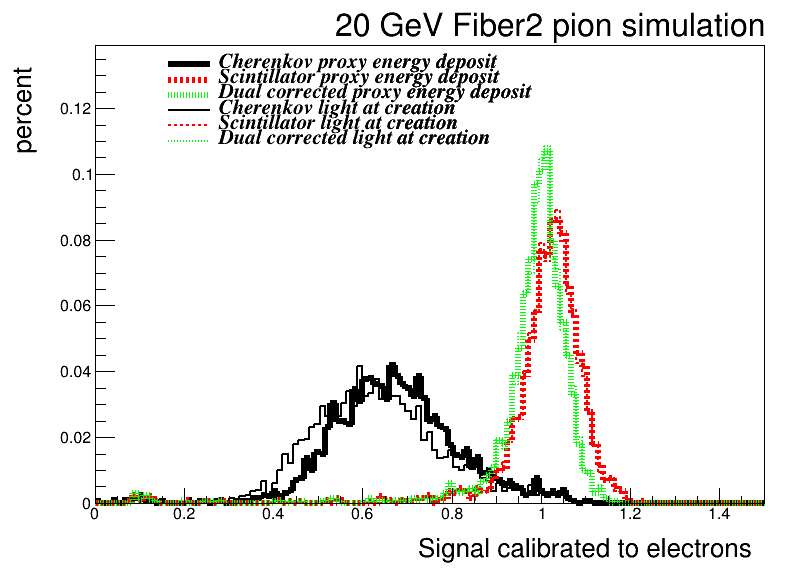}}
\caption{ 
Scintillation, Cherenkov, and dual-readout corrected energy distributions for calorimeters
[top] PbWO,
[middle] Fiber1, Fiber2,
[bottom] SampL, SampS.
The thick lines represent signals based on energy deposits while the thin lines are based on counts of created photons (without propagation to photodetectors).
 }\label{fig:threeres}.  
\end{figure}

The distributions of $S$, $C$, and $D$ are shown in Fig.~\ref{fig:threeres}.  For some calorimeters (PbWO, SampL, Fiber1), the dual readout correction markedly improves the resolution.  For  Fiber2, the improvement is modest.  For calorimeter SampS, the dual-readout corrected energy has worse resolution.

\begin{figure}[hbtp]
\centering
\resizebox{0.480\textwidth}{!}{\includegraphics{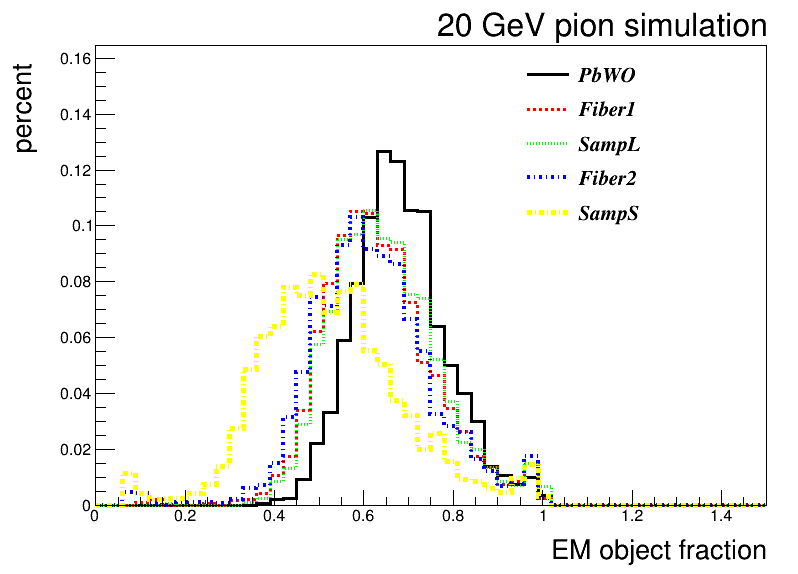}}
\caption{ 
Shower EM object fraction for pion showers for the various calorimeters.
}\label{fig:pfff}.  
\end{figure}

\begin{figure}[hbtp]
\centering
\resizebox{0.480\textwidth}{!}{\includegraphics{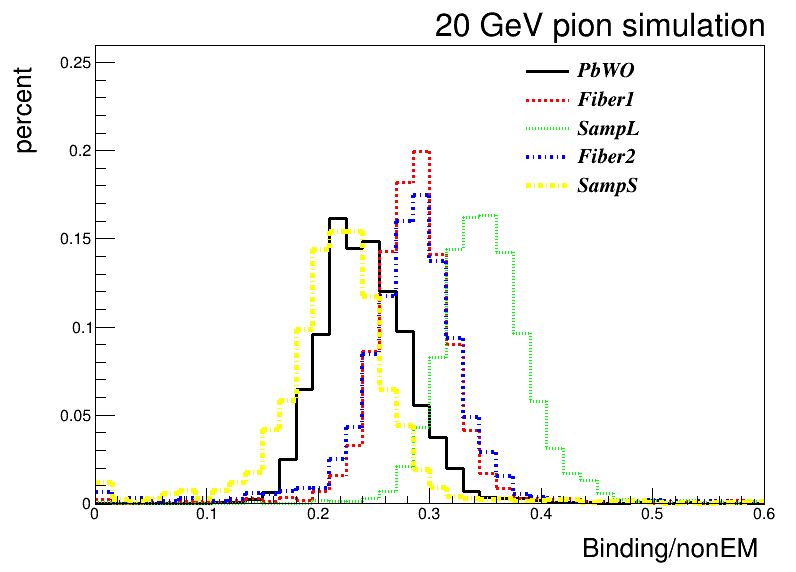}}
\resizebox{0.480\textwidth}{!}{\includegraphics{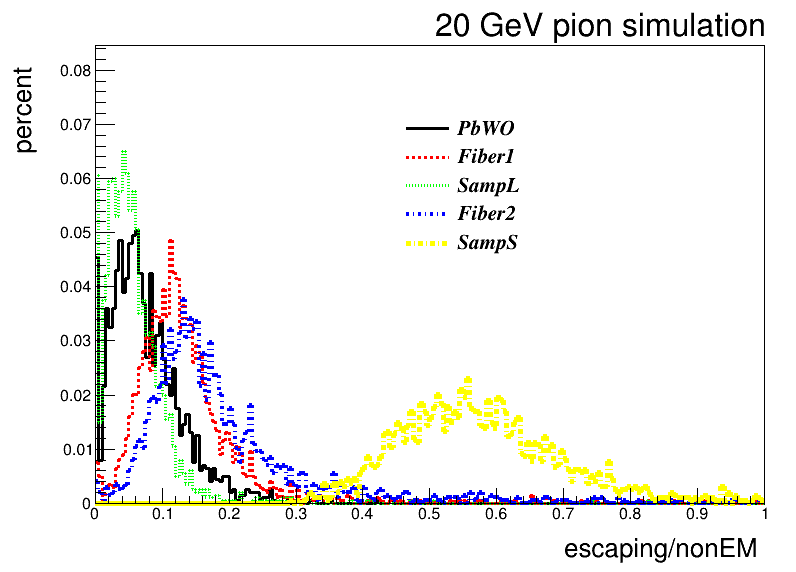}}
\caption{ 
[left] Total  energy loss to binding energies etc. for calorimeters divided by the non-EM energy  
[right] Escaping energy divided by the non-EM energy 
}\label{fig:nonCons}.  
\end{figure}

Figure~\ref{fig:pfff} shows the EM object fraction of the pion shower.  Note these distributions have only a weak dependence on the calorimeter type.  For all except SampS, they peak at 0.6. They are roughly Gaussian in shape.

Figure~\ref{fig:nonCons} [left] shows the total energy loss to binding energies {\it etc}. divided by the non-EM energy for the calorimeters.  Typically about 30\% of the non-EM energy is lost due to mechanisms that do not produce signal, and it is roughly independent of the calorimeter type.  The distribution is roughly Gaussian.

Figure~\ref{fig:nonCons} [right] shows the total energy escaping the calorimeter divided by the non-EM energy for the calorimeters.  
Note that calorimeters PbWO and SampL have resolutions dominated by energy losses due to binding energy, while that of SampS is dominated by escaping energy due to its small transverse dimensions.  
Calorimeters Fiber1 and Fiber2 have similar contributions from both.

\begin{figure}[hbtp]
\centering
\resizebox{0.48\textwidth}{!}{\includegraphics{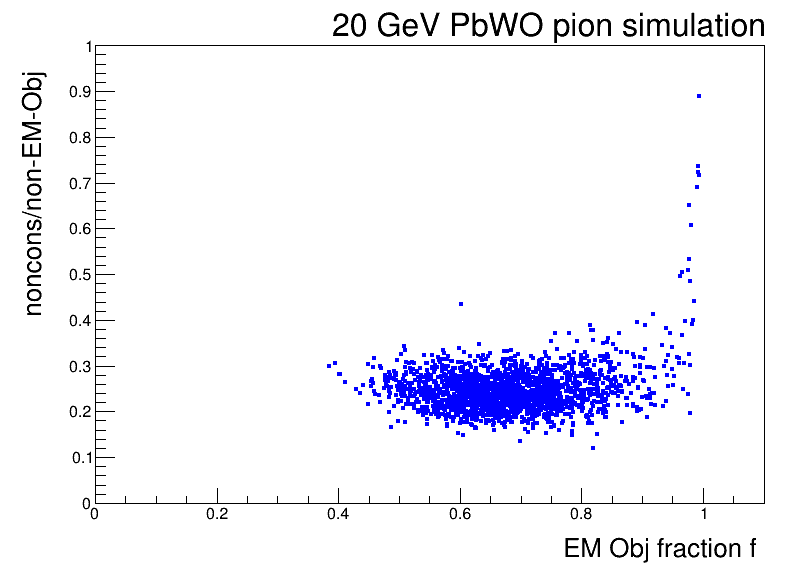}}
\resizebox{0.48\textwidth}{!}{\includegraphics{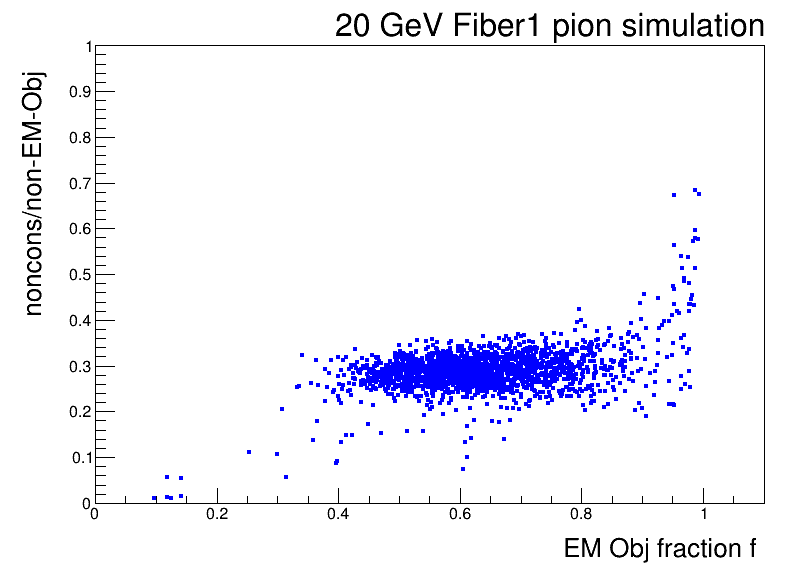}}
\resizebox{0.48\textwidth}{!}{\includegraphics{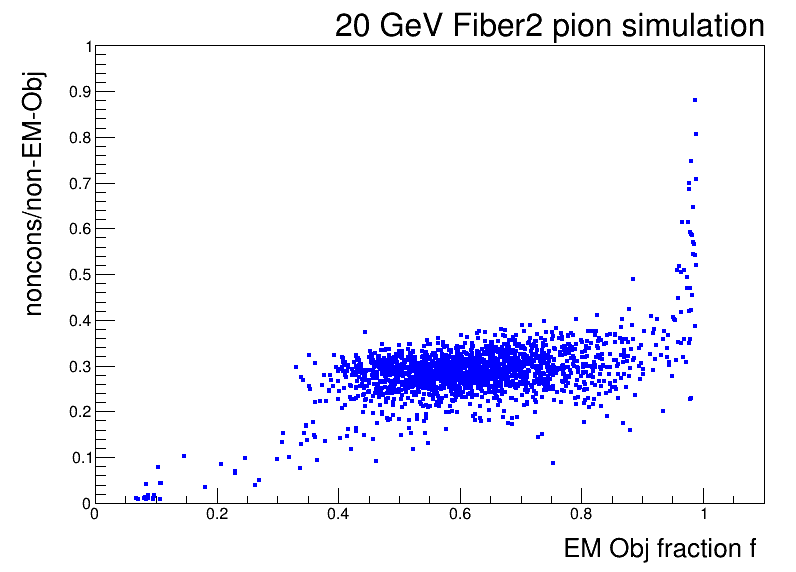}}
\resizebox{0.48\textwidth}{!}{\includegraphics{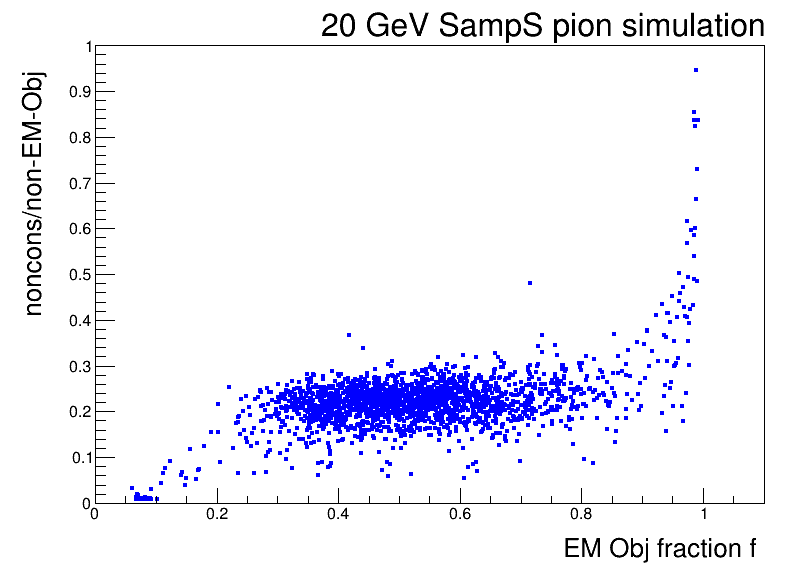}}
\resizebox{0.48\textwidth}{!}{\includegraphics{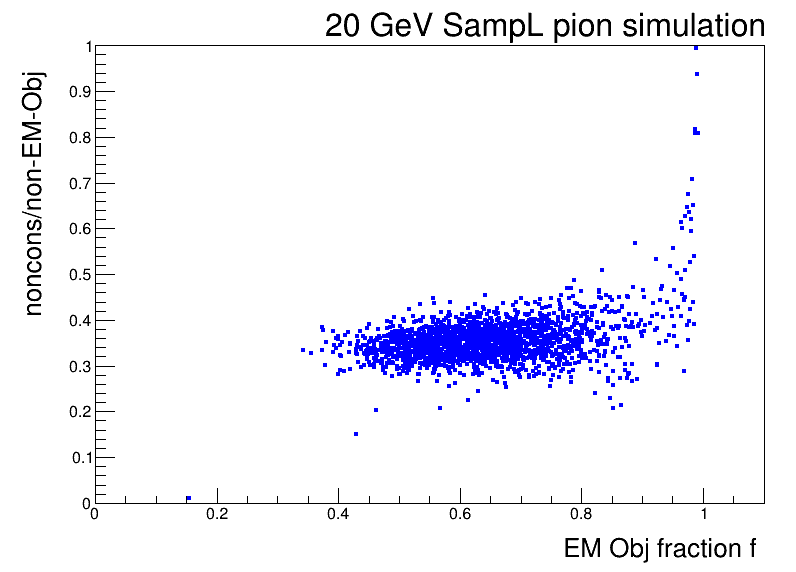}}
\caption{ The fraction of non-EM energy utilized for overcoming nuclear binding energies $f_{NC}$
for [top row] PbWO,
[middle row] Fiber1, Fiber 2
and [bottom] SampS, SampL calorimeters.
 }\label{fig:vf1}.  
\end{figure}

\begin{figure}[hbtp]
\centering
\resizebox{0.48\textwidth}{!}{\includegraphics{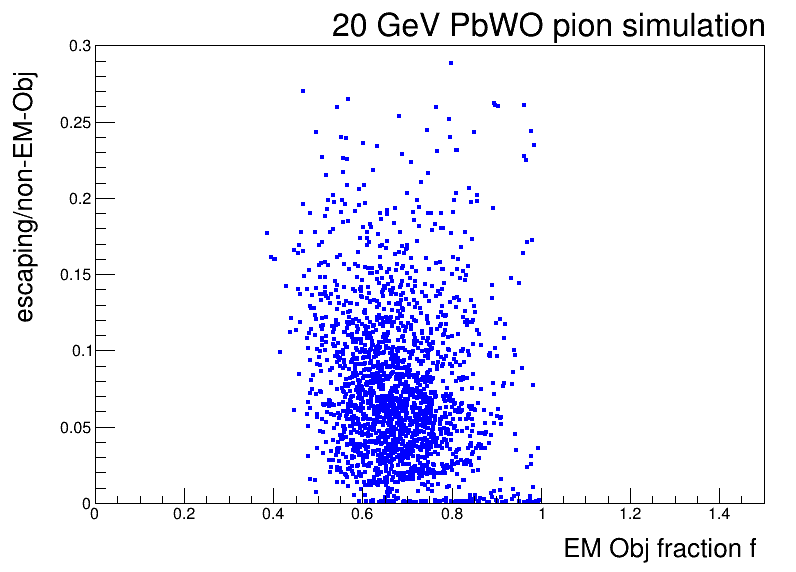}}
\resizebox{0.48\textwidth}{!}{\includegraphics{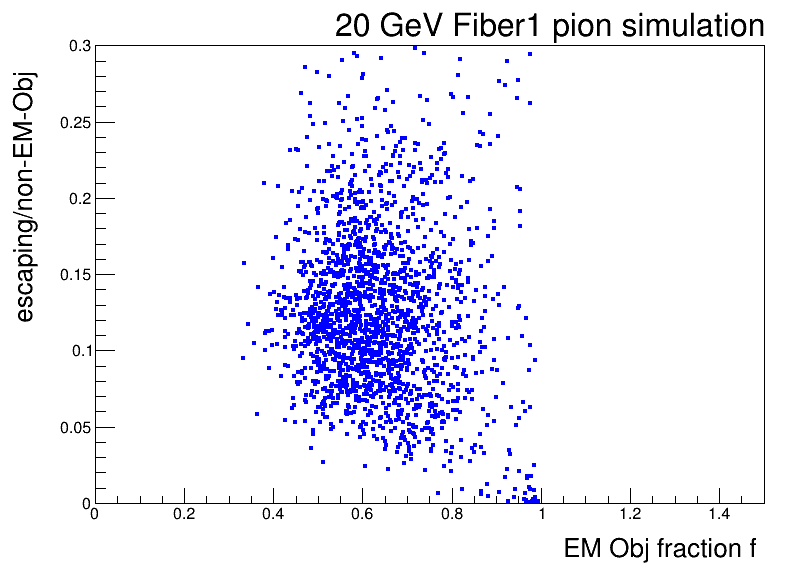}}
\resizebox{0.48\textwidth}{!}{\includegraphics{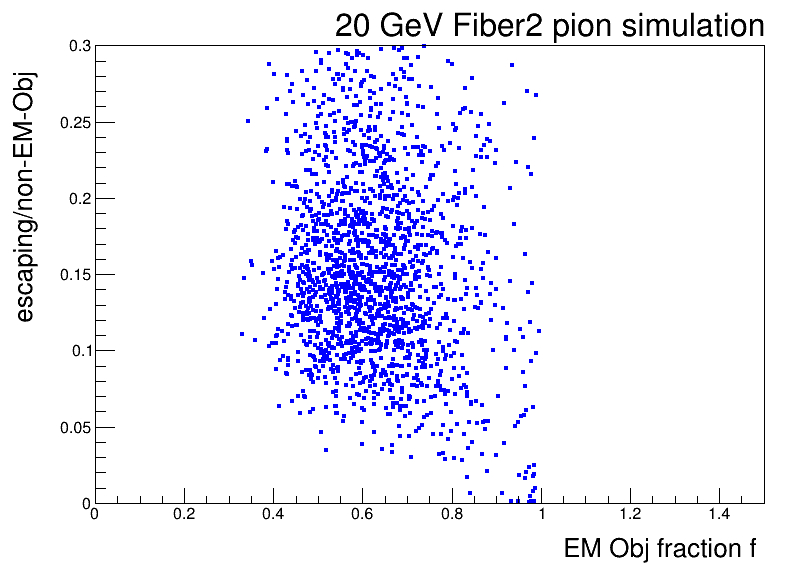}}
\resizebox{0.48\textwidth}{!}{\includegraphics{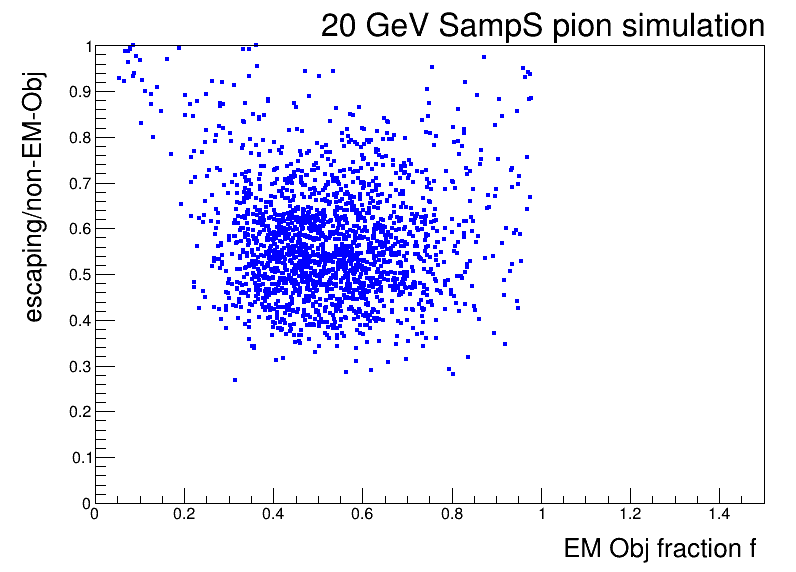}}
\resizebox{0.48\textwidth}{!}{\includegraphics{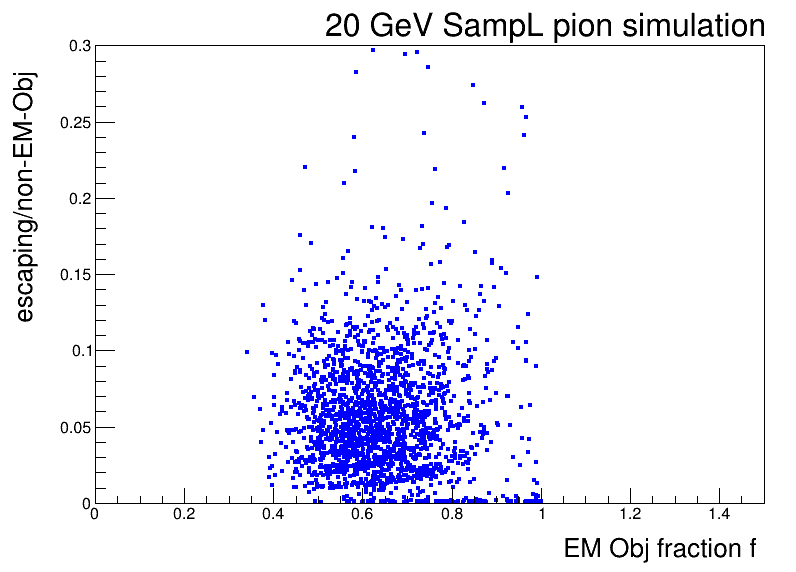}}
\caption{The
fraction of non-EM energy escaping the calorimeter $f_{ES}$  
for [top row] PbWO,
[middle row] Fiber1, Fiber2,
and [bottom row] SampS, SampL calorimeters.
 }\label{fig:vf2}.  
\end{figure}

\begin{figure}[hbtp]
\centering
\resizebox{0.550\textwidth}{!}{\includegraphics{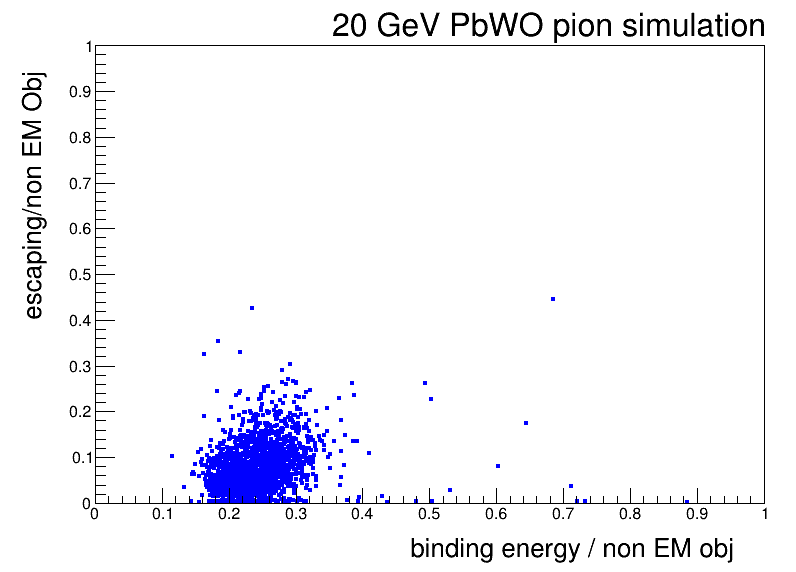}}
\resizebox{0.480\textwidth}{!}{\includegraphics{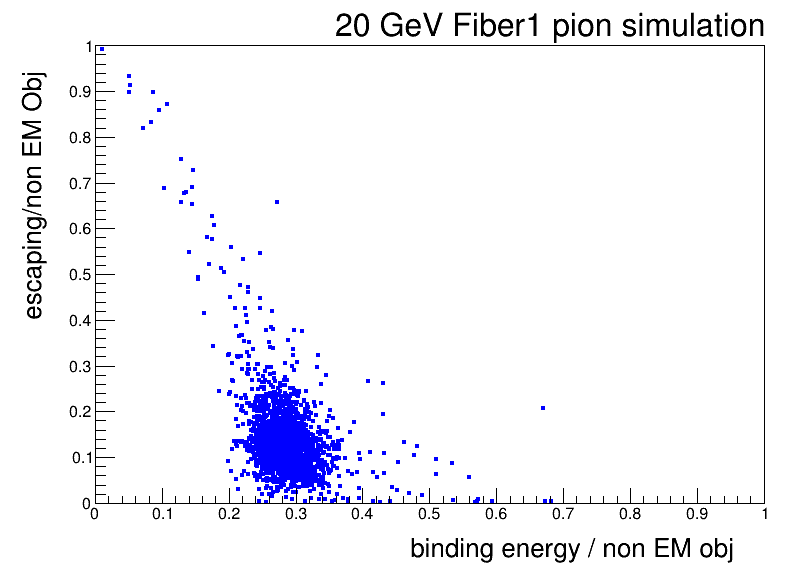}}
\resizebox{0.480\textwidth}{!}{\includegraphics{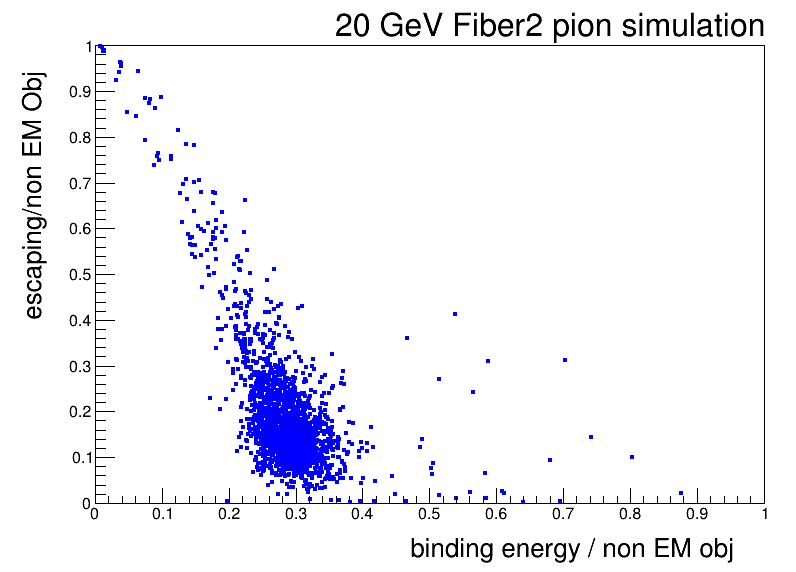}}
\resizebox{0.480\textwidth}{!}{\includegraphics{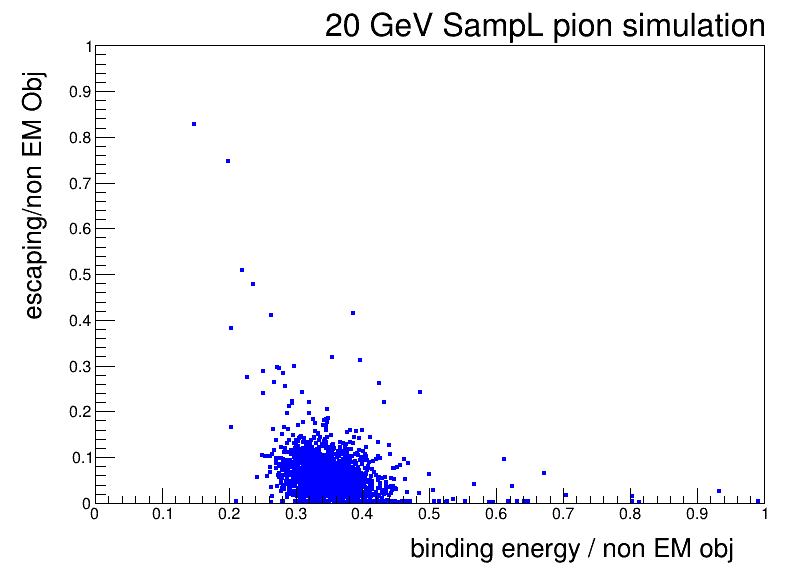}}
\resizebox{0.480\textwidth}{!}{\includegraphics{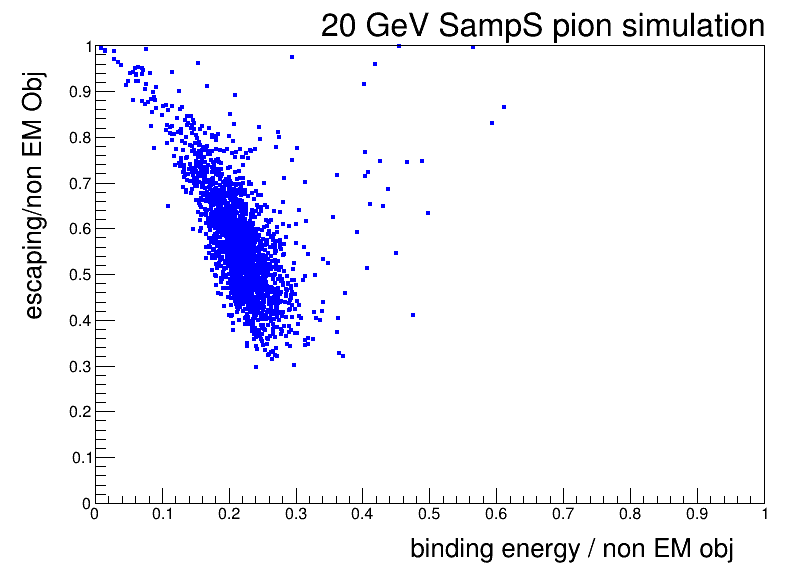}}
\caption{ [left] The fraction of non-EM energy going to overcoming  nuclear binding energies $f_{NC}$ versus
fraction of non-E energy escaping the calorimeter $f_{ES}$ [right] 
for [top row] PbWO,
[middle row] Fiber1, Fiber2,
and [bottom row] SampS, SampL calorimeters.
 }\label{fig:escvnon}.  
\end{figure}

Figures~\ref{fig:vf1} and \ref{fig:vf2} show $f_{NC}$ and $f_{ESC}$ versus $f$ respectively for the various calorimeters.  The ansatz of Eq.~\ref{eqn:ansatz2} only works if these variables are uncorrelated.  This assumption is reasonable (although not perfect).
The ansatz also only works if $f_{ES}$ and $f_{NC}$ are not correlated with each other.  Their correlations are shown in Fig.~\ref{fig:escvnon}. The assumption is strongly violated for calorimeter SampS.  It is most true for calorimeters PbWO and SampL.  This does not affect the prediction for the improvement when moving to a dual-readout corrected energy, but does affect the simple formula for $h_s$ given in Eq.~\ref{eqn:hs2}.

\begin{table}[!htp]
\centering
\caption{Sampling fractions for electrons and pions for scintillation and Cherenkov signals.}
\begin{tabular}{lcccccc}
\hline
label & $g_e$  & $g_\pi(f=0)$  & $g_\pi(f=1)$ & 
$g_e$  & $g_\pi(f=0)$  & $g_\pi(f=1)$  \\ 
 & scint. &  scint. &  scint.& 
 cher. &  cher. & cher.  \\ \hline
Fiber1 &  0.02521 & 0.04855 & 0.02319 & 
0.03928 & 0.05472 & 0.04201 \\
      & $\pm$ 0.00006 & $\pm$ 0.0003  & $\pm$ 0.0007 & 
      $\pm$ 0.00005 &$\pm$ 0.0004 & $\pm$  0.0007\\
Fiber2 &  0.05469 & 0.09177 & 0.05294 & 
0.09312 & 0.1204 &  0.09686\\
      & $\pm$ 0.00005 & $\pm$ 0.0005  & $\pm$  0.0008  &
      $\pm$ 0.00009 &$\pm$  0.0005& $\pm$  0.0008\\
SampL &  0.03414 & 0.0671 & 0.0339& 
0.06237 & 0.07827 & 0.0629 \\
      & $\pm$ 0.00005 & $\pm$  0.0008 & $\pm$ 0.001 &
      $\pm$ 0.00008 &$\pm$ 0.0008  & $\pm$ 0.001 \\
SampS &  0.03512 & 0.04819 & 0.03297 & 
0.03413 & 0.04857 & 0.03121 \\
      & $\pm$ 0.00004 & $\pm$ 0.003 & $\pm$ 0.006 & 
      $\pm$ 0.00003& $\pm$ 0.004 & $\pm$ 0.006 \\
\hline
\end{tabular}
\label{tab:calibc}
\end{table}

\subsection{Simulation results}\label{sec:results}

To calculate the predicted resolution using Eq.~\ref{eqn:maineqn}, we need $f_{res}$, $h_S$, and $h_C$.
We obtain  $<f>$ and $f_{res}$  from  Fig~\ref{fig:pfff} for each geometry. Resolutions are the rms of a fit to a Gaussian near the most probable value.
The values of $h_S$ and $h_C$ are calculated as:
\begin{equation}
    h_S=\frac{<S>-<f>}{1-<f>}
    \label{eqn:hs1}
\end{equation}
and
\begin{equation}
    h_C=\frac{<C>-<f>}{1-<f>}.
\end{equation}
Using these definitions, Eq.~\ref{eqn:maineqn}
becomes
\begin{equation}
    \sigma_D = \frac{\sqrt{(1-<C>)^2 \sigma_S^2 + (1-<S>)^2 \sigma_C^2 -2 \frac{(1-<C>)^2(1-<S)^2}{(1-<f>)^2}f_{res}^2}}{<S>-<C>}.
\end{equation}

The measured values for the various calorimeters are given in Tables~\ref{tab:averagevalues} and ~\ref{tab:rmsvalues}.
Note that for calorimeter SampL, the dual readout correction did not yield a response of one, as by definition it should.  Because of the small size and large fluctuations for this calorimeter, it is challenging to calibrate.
For SampS, our values are $\sigma_S/S=$ 0.159$\pm$0.006, $\sigma_C/C=$0.279$\pm$0.014, and $\sigma_D/D=$0.23$\pm$0.02. 
 The values from Ref.~\cite{chekanov2023geant4simulationssamplinghomogeneous} are
  $0.196\pm0.004$, $0.293\pm0.006$,and $0.232\pm 0.005$.
 Note that the simulation in Ref.~\cite{chekanov2023geant4simulationssamplinghomogeneous} did photon creation and propagation, and so includes photo statistics, while our simulation does not.  Both simulations however show the dual-readout correction causing worse resolution.
 For calorimeter Fiber1, we get a dual-readout-corrected resolution of 0.055$\pm$0.004.
 In Ref.~\cite{Lucchini:2020bac}, read from Fig.~22, the resolution is the same to within our ability to read the graph.
Using these values and the formula, the predicted dual readout resolution is given in Table~\ref{tab:compare}.

\begin{table}[!htp]
\centering
\caption{Average values for the scintillation S, Cherenkov C, and Dual-Readout corrected D energy estimates.
Uncertainties are statistical.}
\begin{tabular}{lllll }
\hline
label & $<f>$ & $<S>$ &  $<C>$ & $<D>$  \\
         \hline

PbWO    & 0.669     & 0.914     & 0.671     & 1.002 \\
        &$\pm$0.004 &$\pm$ 0.001&$\pm$0.003 &$\pm$ 0.001\\
 Fiber1 &0.606      & 1.076     & 0.690     & 0.998 \\
        &$\pm$0.005 &$\pm$0.003 &$\pm$0.006 &$\pm$ 0.003 \\
Fiber2  & 0.599     & 1.03      & 0.657      & 1.00 \\
        &$\pm$0.006 &$\pm$0.002 &$\pm$0.006 &$\pm$ 0.002\\
SampL   & 0.626     & 1.108     & 0.654      & 1.01  \\
        &$\pm$0.005 &$\pm$0.004 &$\pm$ 0.004 &$\pm$ 0.003 \\
SampS   & 0.503     & 0.725     & 0.521     & 0.94\\
        &$\pm$0.007 &$\pm$0.004 &$\pm$0.007 &$\pm$0.01 \\
\hline
\end{tabular}
\label{tab:averagevalues}
\end{table}

\begin{table}[!htp]
\centering
\caption{Resolution values  for the scintillation S, Cherenkov C, and Dual-Readout corrected D energy estimates.
Uncertainties are statistical.  }
\begin{tabular}{lllll }
\hline
label &  $\sigma_f$ &  $\sigma_S$ &  $\sigma_C$ & $\sigma_D$  \\ \hline
PbWO   & 0.086     & 0.031     & 0.086     & 0.028 \\
       &$\pm$0.005 &$\pm$0.002 &$\pm$0.005 &$\pm$0.002 \\
Fiber1 & 0.109     & 0.081     & 0.123     &0.055\\
       &$\pm$0.007 &$\pm$0.005 &$\pm$0.009 & $\pm$0.004\\
Fiber2 & 0.12      & 0.048     & 0.124     & 0.041\\
       &$\pm$ 0.1   &$\pm$0.003 &$\pm$0.009 & $\pm$ 0.002\\
SampL  & 0.116      & 0.098     & 0.102     &0.067 \\
       &$\pm$0.009 &$\pm$ 0.006&$\pm$0.006 &$\pm$ 0.004\\
SampS  & 0.15      &$\pm$0.115  & 0.15     &0.22 \\
      &$\pm$ 0.01 &$\pm$ 0.006  &$\pm$ 0.01 &$\pm$ 0.02  \\
\hline
\end{tabular}
\label{tab:rmsvalues}
\end{table}

\begin{table}[!htp]
\centering
\caption{Comparison between the dual-readout corrected energy estimate resolution predicted using Eq.~\ref{eqn:maineqn} and  the measured value given in Table~\ref{tab:rmsvalues}.  The first column represents the dual-readout corrected resolution using Eq.~\ref{eqn:maineqn}. The second column is the difference between this number and the one in Tab.~\ref{tab:rmsvalues}, divided by the uncertainty.  
Uncertainties are statistical. }
\begin{tabular}{lcc}
\hline
PbWO &0.027 $\pm$ 0.006      & 0.1\\
Fiber1 & 0.065 $\pm$  0.004  & -1.8 \\
Fiber2 & 0.043 $\pm$  0.003  &-0.8\\
SampL & 0.070 $\pm$ 0.006    &-0.5\\
SampS & 0.19 $\pm$ 0.04      & 0.4\\

\hline
\end{tabular}
\label{tab:compare}
\end{table}

In addition, we can predict $h_S$ using either Eq.~\ref{eqn:hs2} or  Eq.~\ref{eqn:hs1}.  The results are in given Table~\ref{tab:hs}.
The agreement is good except for calorimeter SampL, which has strong correlations between $f_{ES}$ and $f_{NC}$, rendering Eq.~\ref{eqn:hs2} invalid.

\begin{table}[!htp]
\centering
\caption{The value of $h_S$ calculated using Eq.~\ref{eqn:hs2} and  Eq.~\ref{eqn:hs1}.
For the Eq.~\ref{eqn:hs2} calculation, values are needed for $<f_{NC}>$ (column 2), 
$<f_{ES}>$ (column 3), the ratio of the non-EM to EM sampling fractions (column 3, for PbWO by definition this is one).  
Column 4 gives the $h_S$ value from  Eq.~\ref{eqn:hs2}.
Column 5 gives the value from Eq.~\ref{eqn:hs1}.
The last column gives their ratio.
  }
\begin{tabular}{lcccccc}
\hline
name &$<f_{NC}>$ & $f_{ESC}$ & g ratio & $h_S$ Eq.~\ref{eqn:hs2} & $h_S$ Eq.~\ref{eqn:hs1}  &pred/meas  \\ \hline
PbWO   & 0.238       & 0.062         & 1            & 0.70          & 0.740       & 0.95\\
       & $\pm$ 0.001 & $\pm$ 0.001   & $\pm$ 0.000  & $\pm$ 0.02    & $\pm$0.003      & $\pm$ 0.02 \\
Fiber1 & 0.287       & 0.121         & 2.09         & 1.24          & 1.192        & 1.04\\
       & $\pm$ 0.001 & $\pm$ 0.001   & $\pm$ 0.06   & $\pm$ 0.04    & $\pm$0.009      & $\pm$ 0.03 \\
Fiber2 & 0.287      & 0.152         &  1.73         & 0.97          & 1.075       &0.91\\
       & $\pm$ 0.001 & $\pm$ 0.002   & $\pm$ 0.03   & $\pm$ 0.02    & $\pm$0.005      & $\pm$0.02 \\
SampL  & 0.347      & 0.046          & 1.98         & 1.20          & 1.29        & 0.93\\
       & $\pm$ 0.001 & $\pm$ 0.001   & $\pm$ 0.08   & $\pm$ 0.06    & $\pm$0.01      & $\pm$ 0.05 \\
SampS  & 0.219      & 0.549          & 1.46         & 0.34          & 0.45       & 0.75\\
       & $\pm$ 0.001 & $\pm$ 0.004   & $\pm$ 0.28   & $\pm$  0.06   & $\pm$0.01      & $\pm$ 0.15 \\
\hline
\end{tabular}
\label{tab:hs}
\end{table}

In addition, we can predict the $S$ resolution using Eq.~\ref{eqn:scintrespred}.
This formula is only valid if $g_2$, $f$, $f_{SC}$, and $f_{NC}$ are uncorrelated. It would
be surprising if $g_2$ and $f_{NC}$ are uncorrelated.
The results are shown in Tab.~\ref{tab:sresp}.  The agreement is reasonable, at the 10\% level, except for Fiber2, despite the violation of the conditions.

\begin{table}[!htp]
\centering
\caption{
The value of $\sigma_s/<S>$ calculated using Eq.~\ref{eqn:scintrespred} compared to the direct value from Tab.~\ref{tab:rmsvalues}.
For the Eq.~\ref{eqn:scintrespred} calculation, values are needed for $\sigma_f$ (column 2) and
$\sigma_g/g$ (column 3). By definition, $\sigma_g/g$ for PbWO4 is zero.
Column 4 gives $\sigma_s/<S>$ calculated using Eq.~\ref{eqn:scintrespred} 
Column 5 gives the ratio of column 4 to the $\sigma_S$ value in Tab.\ref{tab:rmsvalues}
The last column gives their ratio.
}
\begin{tabular}{lccccc}
\hline
name &$\sigma_f$ & $\sigma_g/g$  & $\sigma_s/<S>$  Eq.~\ref{eqn:scintrespred}   & ratio\\ \hline
PbWO    & 0.0886      &  0            & 0.0332        & 0.973 \\
        & $\pm$ 0.005 & $\pm$0.0      & $\pm$0.00002  &  $\pm$0.05      \\
Fiber1  & 0.109       & 0.0515        & 0.07135       &  0.948 \\
        & $\pm$ 0.007 & $\pm$0.005    & $\pm$0.0001   & $\pm$ 0.06       \\
Fiber2  &  0.12       & 0.033         & 0.0701        &  1.51 \\
        & $\pm$ 0.01  & $\pm$0.003    & $\pm$0.0002   & $\pm$ 0.09       \\
SampL   &0.116        & 0.074         & 0.0894        &  1.01 \\
        & $\pm$ 0.009 & $\pm$0.007    & $\pm$ 0.0001  & $\pm$0.07       \\
SampS   &0.15         & 0.12          & 0.1785        &  1.12 \\
        & $\pm$ 0.01  & $\pm$0.02     &  $\pm$ 0.003  & $\pm$ 0.06       \\
\hline
\end{tabular}
\label{tab:sresp}
\end{table}


\section{Conclusions}
We have presented a simple formula that predicts the improvement (or lack thereof) when utilizing the dual readout energy estimate for a calorimeter. The formula works for a wide variety of calorimeters, including both calorimeters whose response variation is dominated by losses to overcoming nuclear binding energy and those dominated by fluctuations in energy escaping the calorimeter.  Formulae are also presented for estimating the scintillation energy scale and resolution in terms of the sampling fraction, fraction of escaping energy, fraction of the shower that is EM objects, and fraction of energy lost to overcoming binding energies.  These results should help intuition on the impact of the dual readout correction for a wide variety of calorimeter geometries.

\section*{Acknowledgments}
This work was supported in part by U.S. Department of Energy Grant  DE-SC0022045. 
Argonne National Laboratory’s work was funded by the U.S. Department of Energy, Office of High
Energy Physics under contract DE-AC02-06CH11357.
We thank Bob Hirosky for his comments on the manuscript.


\bibliographystyle{elsarticle-num}
\bibliography{main}
\newpage
\appendix
\section{Derivation of the Covariance}
\label{app:app1}

In this appendix, we derive the covariance between $S$ and $C$. The following notations are introduced for clarity

\begin{equation}
    H \equiv (1-h_S)(1-h_C),
\end{equation}
where $h_S$ is the average response to the non-EM part of the shower for the $S$ measurement, and $h_C$ is the corresponding response for the $C$ measurement.

\begin{equation}
    F_p \equiv \,<f>^2 + \sigma_{f}^2,
\end{equation}
\begin{equation}
    F_m \equiv  \sigma_{f} <f>,
\end{equation}
where $\sigma_f$ is the rms of the Gaussian fluctuating variable $g_1$, representing fluctuations in the EM portion of the shower, and $<f>$ is its mean value.

\begin{equation}
    Q_i \equiv \left(\frac{1}{\sqrt{2\pi}}\right)^i \frac{1}{\sigma_{f} n_S n_C}, \quad i=0,1,2,3.
\end{equation}
where $n_S$ and $n_C$ are the rms values of $g_2$ and $g_3$, which represent the noise (or other non-correlated resolution terms) in the $S$ and $C$ measurements.

To calculate $cov(S,C)$, we start with the integral part of Eq.\ref{eqn:eq9}

\begin{equation}
\begin{aligned}
    & \int_{-\inf}^{\inf}  \int_{-\inf}^{\inf}  \int_{-\inf}^{\inf} 
    \big( (x_1 + (1-x_1)h_S) + x_2 \big) 
    \big( (x_1 + (1-x_1)h_C) + x_3 \big)
    g_1 g_2 g_3 \, dx_1 dx_2 dx_3 \\
    &= Q_3 \iiint
    \big[ (1-h_S)x_1 + x_2 + h_S \big] 
    \big[ (1-h_C)x_1 + x_3 + h_C \big]
    \exp\left(-\frac{(x_1 - <f>)^2}{2\sigma_{f}^2}\right) \\
    & \quad \times \exp\left(-\frac{x_2^2}{2n_S^2}\right) 
    \exp\left(-\frac{x_3^2}{2n_C^2}\right) \, dx_1 dx_2 dx_3\\
    &= Q_2\iiint
\big[HF_p\sigma_{f} + (x_3+h_C)(1-h_S) F_m +(x_2+h_S)(1-h_C)F_m+ (x_2+h_S)(x_3+h_C)\sigma_{f}\big]\\
& \quad \times \exp\left(-\frac{x_2^2}{2n_S^2}\right)\exp\left(-\frac{x_3^2}{2n_C^2}\right)dx_2 dx_3\\
&=Q_1\iiint
\big[ HF_p\sigma_{f}n_S + (x_3+h_C)(1-h_S) F_m n_S +(h_S-h_Ch_S)F_mn_S+ h_S(x_3+h_C)\sigma_{f}n_S\big]\\
& \quad \times \exp\left(-\frac{x_3^2}{2n_C^2}\right) dx_3\\
&=Q_0
\big[ HF_p\sigma_{f}n_S + h_C(1-h_S) F_m n_S +(h_S-h_Ch_S)F_mn_S+h_Sh_C\sigma_{f}n_S\big]n_C \\
&= HF_p + h_C(1-h_S) <f> +h_S(1-h_C)<f>+h_Sh_C.
\end{aligned}
\end{equation}

According to Eq.\ref{eqn:eq9}, the covariance between $S$ and $C$ is then given by

\begin{equation}
\begin{aligned}
    cov(S,C) &= \left[ H F_p + h_C(1-h_S)<f>+ h_S(1-h_C)<f> + h_S h_C \right] \\
    & \quad - (<f>+ (1-<f>)h_S)(<f> + (1-<f>)h_C) \\
    &= H F_p - H <f>^2 \\
    &= H \sigma_{f}^2.
\end{aligned}
\end{equation}

This result demonstrates that the covariance depends on the product of $H$ and the variance $\sigma_{f}^2$.

\end{document}